\documentclass[trackchanges,twocolumn]{aastex7}

\usepackage{amsmath}
\usepackage{xcolor}
\usepackage{textcomp}
\usepackage{mathrsfs}
\usepackage{gensymb}

\shorttitle{Reconciling sGRB and GW neutron star merger rates}
\shortauthors{Kunnumkai et al.}
\received{\today}
\revised{}
\accepted{}

\submitjournal{ApJ}
\graphicspath{{./}{Figures/}}
\usepackage{hyperref} 
\begin{document}

\title{Wide Jets or Low Rates: Reconciling Short GRB and Gravitational-Wave Neutron Star Merger Rates}

\author[0009-0000-4830-1484]{Keerthi Kunnumkai}
    \affiliation{McWilliams Center for Cosmology and Astrophysics, Department of Physics, Carnegie Mellon University, Pittsburgh, PA 15213, USA}
    \email[show]{kkunnumk@andrew.cmu.edu}  
\author[0000-0002-6011-0530]{Antonella Palmese}
    \affiliation{McWilliams Center for Cosmology and Astrophysics, Department of Physics, Carnegie Mellon University, Pittsburgh, PA 15213, USA}
    \email{apalmese@andrew.cmu.edu}
\author[0000-0002-9700-0036]{Brendan O'Connor}
    \altaffiliation{McWilliams Fellow}
    \affiliation{McWilliams Center for Cosmology and Astrophysics, Department of Physics, Carnegie Mellon University, Pittsburgh, PA 15213, USA}
    \email{boconno2@andrew.cmu.edu} 
\author[0000-0002-6121-0285]{Amanda Farah}
    \affiliation{Canadian Institute for Theoretical Astrophysics, University of Toronto, 60 St George St,  Toronto, ON M5S 3H8, Canada}
     \email{afarah@cita.utoronto.ca}
\author[0000-0003-2362-0459]{Ignacio Maga\~na~Hernandez}
    \altaffiliation{McWilliams Fellow}
    \affiliation{McWilliams Center for Cosmology and Astrophysics, Department of Physics, Carnegie Mellon University, Pittsburgh, PA 15213, USA}
    \email{imhernan@andrew.cmu.edu}


\begin{abstract}
Gravitational wave (GW) and short Gamma Ray Burst (sGRB) observations provide us with complementary views of compact object mergers. The paucity of binary neutron star merger (BNS) detections in the latest LIGO/Virgo/KAGRA (LVK) observing run raises the question of whether the GW merger rates are sufficient to explain the observed sGRB rate with compact object mergers alone. We investigate this connection using the latest merger rate constraints from the fourth LVK observing run (O4) and published estimates of the local sGRB rate density. For an observed sGRB rate density of $ \sim 1-7~\mathrm{Gpc^{-3}\,yr^{-1}}$, if $>55\%$ of BNS mergers can successfully launch a jet, we find that the current LVK BNS merger rate can be reconciled with a sGRB merger population containing a significant fraction of relatively wide jets with core half-opening angles $\theta_{\rm j}\gtrsim10\degree$. Meanwhile, a narrow jet population ($\theta_{\rm j}\sim6\degree$) can only be matched with the O4 neutron star merger rate estimates for an observed sGRB rate density of $\lesssim 1~\mathrm{Gpc^{-3}\,yr^{-1}}$, which is consistent with several of the latest estimates. We also find that neutron star-black hole (NSBH) mergers are expected to be a subdominant component of the sGRB population compared to BNS mergers, and they cannot help reconcile some of the highest available sGRB rates ($ >7~\mathrm{Gpc^{-3}\,yr^{-1}}$) with the GW rates. NSBHs can still contribute substantially to the sGRB population, comprising $\sim 6-16\%$ of it for a $\sim 1-3~\mathrm{Gpc^{-3}\,yr^{-1}}$ observed sGRB rate density. Our results indicate that present GW and sGRB observations remain broadly consistent with BNS mergers as the main progenitors of sGRBs.

\end{abstract}


\section{Introduction}
\label{sec:intro}

Short gamma ray bursts (sGRBs) are brief flashes of high energy radiation characterized by prompt emission of less than 2 seconds \citep{1993Kouveliotou}. Since their discovery, compact object mergers have been among the leading progenitor candidates \citep{1984Blinnikov, 1986Paczynski, 1989Eichler, 1992Narayan}. This association was decisively strengthened by the gravitational wave (GW) event GW170817 \citep{2017Abbott,LIGOScientific:2017zic}, which established that at least some binary neutron star (BNS) mergers produce relativistic outflows capable of generating sGRB-like prompt emission \citep{Goldstein_2017, Savchenko_2017}.

Theoretical work has long suggested that neutron star-black hole (NSBH) mergers may also be capable of producing sGRBs, provided the neutron star is tidally disrupted outside the black hole’s innermost stable circular orbit $R_{\rm ISCO}$, leaving sufficient material to form an accretion disk and power a jet \citep{Mochkovitch1993Nature,VossTauris2003MNRAS,Ciolfi2018sGRBReview}. The relative contribution of NSBH systems to the rate of observed sGRBs remains uncertain, as only a subset of NSBH binaries undergo tidal disruption and successfully launch jets \citep{Kyutoku2011PhRvD,ShibataTaniguchi2011LRR,Foucart2012PhRvD} and no multimessenger NSBH detection has yet been reported. 

At the same time, recent observations have revealed that prompt duration alone is not a reliable discriminator of progenitor type. GRB~200826A, despite its short duration, was followed by a broad lined Type Ic supernova \citep{Ahmad2021NatureAstronomy,Rossi_2022}, indicating a collapsar origin \citep{1993Woosley}. Conversely, GRBs~211211A and 230307A exhibited kilonova-like optical emission while having a long duration prompt emission  \citep{2022Natur.612..228T,Rastinejad2022Nature,2022Natur.612..232Y,Levan2023,Yang2024}. While such events highlight the complexity of GRB classification, in this work we restrict our analysis to short duration GRBs and compact object mergers. 

A central quantity in this comparison is the local, on-axis volumetric sGRB rate density, $R_{\rm sGRB}$, defined as the rate of sGRBs whose jets are oriented such that prompt emission would be visible to an on-axis observer, after correcting for instrumental selection (flux threshold, triggering, and sky exposure), but prior to geometric beaming corrections. Comparisons between sGRBs and GW merger rates measured with LIGO/Virgo/KAGRA (LVK) probe both jet collimation and the fraction of mergers that successfully launch a jet. 

In the gravitational wave era, this accounting has become increasingly consequential. While GW170817 provided the definitive multimessenger link between BNS mergers and sGRBs \citep{LIGOScientific:2017zic}, only one additional high-significance BNS candidate (GW190425; \citealt{GW190425}) has been reported to date, and the continued absence of new BNS detections despite greatly expanded observing time has driven the inferred local BNS merger rate substantially downward \citep{GW190425,KAGRA:2021vkt,gwtc4,gwtc4pop}. Meanwhile, detections of NSBH  mergers including black holes in the lower mass gap ($\sim 3-5~M_\odot$), such as GW230529 \citep{LIGOScientific:2024elc} and potentially the candidate S250206dm \citep{GCN250818kLVK}, are expected to be promising multimessenger sources \citep{230529_everything,Kunnumkai:2024qmw,Kunnumkai:2024tuq} that could contribute to the sGRB population \citep{2026Kaur}, although no electromagnetic counterpart has yet been detected \citep{230529_GRB,Hu_2025}.
\begin{figure*}[!ht]
    \includegraphics[scale=0.81]{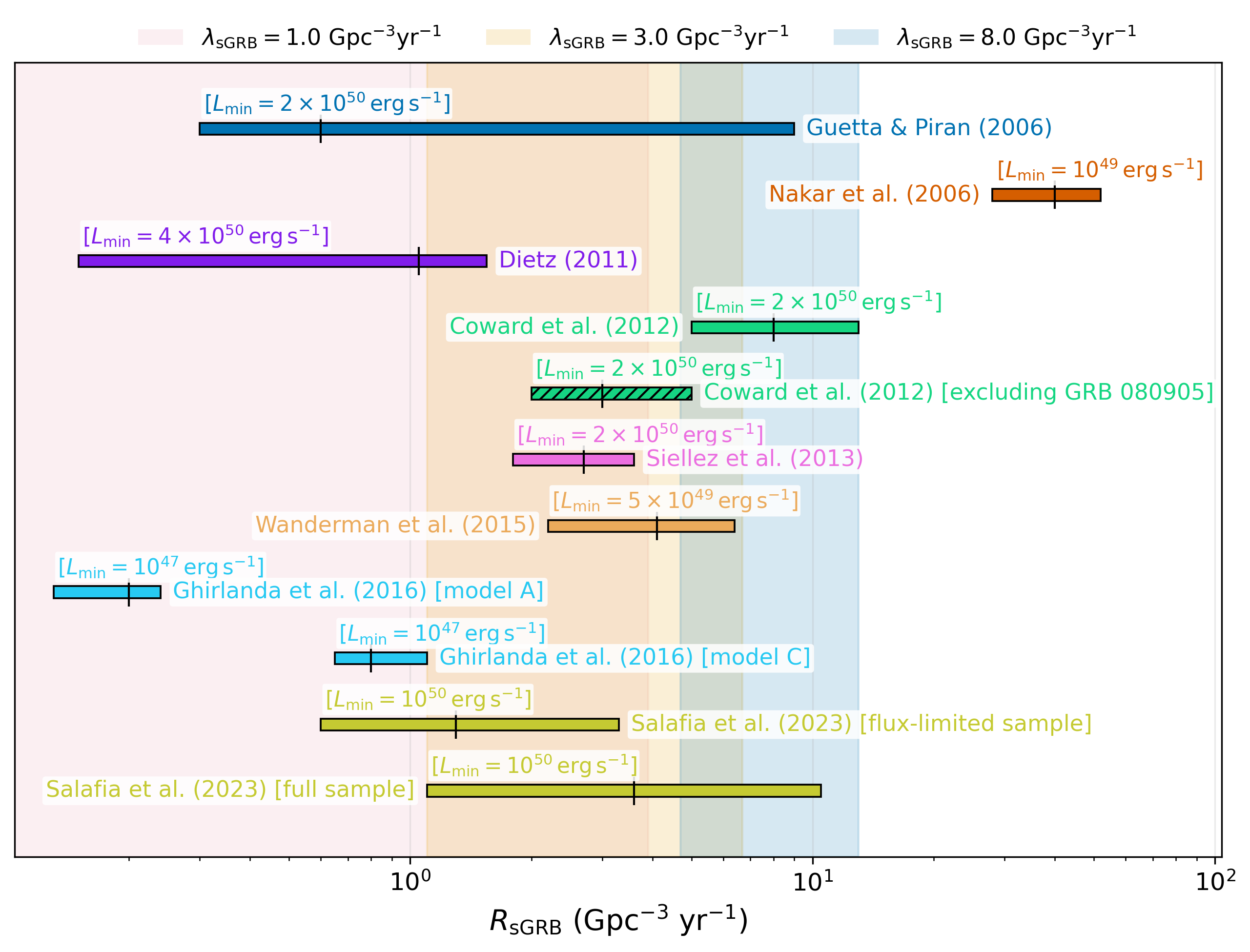}
    \caption{Compilation of local sGRB rate density estimates from the literature. Each horizontal bar shows the 90\% credible interval reported by the corresponding study, while the black vertical line marks the median value. The inferred rate depends  on the sample selection as well as on assumptions on the minimum luminosity ($L_{\rm min}$), as indicated alongside each estimate. The shaded intervals corresponds to 90\% credible intervals of different local sGRB rate intervals (as Poisson distributions with mean $\lambda$) used in this study.}
    \label{fig:rates_lmin}
\end{figure*}

The rapid downward evolution of BNS rate sharpens the potential differences between electromagnetic (EM) inferences of the observed local sGRB rate \citep{Guetta_2007,Nakar2006,Coward2012,Wanderman2015,Ghirlanda2016,2020Dichiara,Salafia2023,Howell2025,pracchia_salafia2026}, and current GW rate constraints \citep{gwtc4,gwtc4pop}. In particular, when plausible beaming-corrections are applied \citep{{Fong2015ApJ815102,2023ApJ...959...13R}}, the intrinsic sGRB rate can be compared with the GW-inferred BNS rate based on jet opening angles and jet-launching fractions \citep{Sarin2022PRD,2023ApJ...959...13R}. Additional uncertainty however arises from the jet's unknown angular structure \citep[e.g,][]{Ryan_2020}, which affects how the beaming correction is interpreted. Although structured emission can broaden the detectable solid angle, particularly in the very local Universe \citep{Howell2025}, sGRBs used to infer the volumetric rate are overwhelmingly detected near on-axis at cosmological distances \citep{OConnor2024MNRAS}. In this regime, the measured on-axis rate is expected to be comparatively insensitive to whether the underlying jets are tophat or structured. The primary impact of jet structure enters through the effective opening angle $\theta_j$ adopted when converting the observed on-axis rate into an intrinsic merger rate.

On the other hand, combining EM and GW measurements offers a powerful way to probe jet formation physics, jet-launching efficiency and the relative role of BNS and NSBH mergers. Understanding how these factors shape the observable sGRB population is central to developing a consistent picture of relativistic jet production in compact mergers.

In this article, we investigate the connection between the observed sGRB rate and the underlying populations of compact object mergers as estimated from the LVK observations from the fourth observing run (\citealt{gwtc4pop}). Building on \citet{Sarin2022PRD}, we incorporate updated merger rate constraints based on GW observations from the first part of the the fourth (O4) observing run (\citealt{gwtc4pop}), and extending those through the end of the O4 run based on the fact there have not been high significance BNS merger alerts. We systematically explore the contributions of BNS and NSBH channels, constraining jet geometry and jet launching efficiency under each scenario, and assess whether compact object mergers alone can explain the observed sGRB rate \citep[see also, e.g.,][]{pracchia_salafia2026,desantis2026,Fishbach_2026}.

In Section \ref{sec:rates} we describe the rates adopted in this work. Section~\ref{sec:methods} introduces the rate framework and modeling assumptions adopted for the plausible sGRB channels. Section~\ref{sec:results} presents the inferred constraints on jet geometry and jet launching efficiency under each progenitor scenario. In Section~\ref{sec:discussion},we examine various systematic uncertainties that could influence the inferred sGRB rates. Finally, Section~\ref{sec:conclusions} summarizes our conclusions.

\section{Rates}
\label{sec:rates}
\subsection{Short Gamma-Ray Bursts}
\label{sec:GRB_rates}

Early estimates of $R_{\rm sGRB}$ were derived by modeling the observed BATSE \citep{BATSE4paper} peak flux distribution. \citet{GuettaPiran2006} assume a broken power-law peak luminosity function and explore different cosmological rate models, including a rate that traces the star formation history and rates delayed relative to it. They conclude that the observed redshift distribution of sGRBs favors a population that lags behind star formation, as expected if sGRBs are associated with compact binary mergers. These preferred models imply $R_{\rm sGRB} = 8\text{--30}~\mathrm{Gpc^{-3}\,yr^{-1}}$, for bursts above $\sim 10^{49}~\rm erg s^{-1}$.

An empirical alternative was presented by \citealt{Coward2012}, where they compute the local sGRB rate density using \emph{Swift} detected bursts with measured redshifts. Instead of using the sGRB luminosity function, they estimate the rate directly from observed bursts within a maximum observable volume $V_{\max}$ \citep{Schmidt1988,Piran1992}, correcting for observational biases. For each burst, they compute the maximum distance and corresponding comoving volume $V_{\max}$ at which \emph{Swift} could detect the event:
\begin{equation}
    V_{\max} = \int_{0}^{z_{\max}} \frac{dV}{dz}\,dz
\end{equation}
The total volumetric rate density is then estimated as a sum over bursts $i$:
\begin{equation}
R_{\rm sGRB}\ \simeq\ \sum_i \frac{1}{V_{{\max},i}}\frac{R_{B/S}}{F_r} \frac{1}{T\Omega} B_i P_i
\label{eq:coward_vmax}
\end{equation}
where ${R_{B/S}} = 6.7$ is the BATSE/\emph{Swift} short GRB ratio, $F_{r}$ accounts for the fact that only a fraction of observed sGRBs have measured redshifts, $B_{i}$ is the beaming fraction, $P_{i}$ is the probability for GRB to be non-collapsar \citep{2013Bromberg}, $T$ is the time span encompassing all \emph{Swift} observations for the sGRB samples they account for, and $\Omega$ = 0.17 is the fractional sky coverage of \emph{Swift}. They derive an observed sGRB rate of $R_{\rm sGRB}$ = $8^{+5}_{-3}~\mathrm{Gpc^{-3}\,yr^{-1}}$ (95\% confidence) assuming isotropic emission, and an upper limit of $\mathfrak{R}$ = $1100^{+700}_{-470}~\mathrm{Gpc^{-3}\,yr^{-1}}$ when corrected for jet-beaming.

\citealt{Wanderman2015} performed a joint analysis of the BATSE, \emph{Fermi} and \emph{Swift} sGRB samples to determine both the luminosity function and rate of non-collapsar sGRBs. In this framework, the BATSE and \emph{Fermi} peak-flux distributions primarily constrain the luminosity function, while the rate is determined primarily by the smaller \emph{Swift} sample that has measured redshifts. To reduce contamination from collapsars in the \emph{Swift} data, they restrict their analysis to bursts with a non-collapsar probability $> 0.6$ \citep{2013Bromberg}, whereas for BATSE and \emph{Fermi} they adopt a sharp duration cut-off of $T_{90} < 2~\rm s$ \citep{1993Kouveliotou}. The luminosity function based approach to deriving the rate is a strong function of the minimum isotropic luminosity of the sample, as there is naturally a higher number of events occurring at increasingly smaller luminosities. For a fiducial $L_{\min}=5\times10^{49}~\mathrm{erg\,s^{-1}}$, \citet{Wanderman2015} infer $R_{\rm sGRB}$ = $4.1^{+2.3}_{-1.9}~\mathrm{Gpc^{-3}\,yr^{-1}}$. They also explicitly demonstrate that the apparent differences between earlier rate estimates (see their Table 4) can be traced to differing implicit assumptions about the low-luminosity cut-off rather than fundamental discrepancies in the observed population (see also Table \ref{fig:rates_lmin}).

\citet{Ghirlanda2016} constrain the sGRB luminosity function, redshift distribution and local rate of sGRBs by fitting a synthetic sGRB population to observational constraints from bursts detected by \emph{Fermi}/GBM and \emph{Swift}. Similar to \citet{Wanderman2015}, they adopt parametric forms of luminosity function and redshift evolution, and extend the analysis by jointly fitting other observables. In particular, they describe the rates derived from two models: In Model A, they assume the existence of intrinsic correlations between peak spectral energy and luminosity (and between peak energy and isotropic energy), while Model C assumes no intrinsic correlation among these quantities. In both cases, the luminosity function is modeled as a broken power law extending down to a minimum luminosity $L_{\min}=5\times10^{47}\,\mathrm{erg\,s^{-1}}$. For Model A, they derive a $R_{\rm sGRB} = 0.20^{+0.04}_{-0.07}~\mathrm{Gpc^{-3}\,yr^{-1}}$. Model C yields a higher rate of $R_{\rm sGRB}= 0.8^{+0.3}_{-0.15}~\mathrm{Gpc^{-3}\,yr^{-1}}$. They attribute this differences to the flatter luminosity function obtained when intrinsic correlations are included, which reduces the number of low-luminosity bursts required to reproduce the observed sample and therefore lowers the inferred rate. Comparing their inferred rate with Galactic BNS merger rate estimates \citep{Kim2015}, \citet{Ghirlanda2016} show that if all BNS mergers produce sGRBs, the implied average jet opening angle must be narrow, $\sim3\degree$-$6\degree$.

\citet{Salafia2023} revisit the sGRB population within a quasi-universal structured jet framework, aiming to reconcile discrepancies between previous works such as \citet{Wanderman2015} and \citet{Ghirlanda2016} by explicitly accounting for viewing angle effects and selection biases. They model the population with parametric distributions for the intrinsic (on-axis) luminosity and redshift evolution but embed these within a universal angular jet structure and perform a hierarchical Bayesian inference. Their analysis combines three datasets: a \emph{Fermi}/GBM sample of sGRBs, a flux-complete \emph{Swift}/BAT sample with associated redshifts, and GW170817 \citep{Abbott_2017grb,Goldstein_2017,Savchenko_2017}. In this framework, the absolute rate is calibrated by requiring the model to reproduce the observed \emph{Fermi}/GBM detection rate, yielding a collimation corrected sGRB rate of $\mathfrak{R} = 740^{+3870}_{-630}~\mathrm{Gpc^{-3}\,yr^{-1}}$ from the full sample and $\mathfrak{R} = 180^{+660}_{-145}~\mathrm{Gpc^{-3}\,yr^{-1}}$ for a flux-limited sample. To enable direct comparison with previous studies, they additionally report beaming uncorrected (on-axis) rates above a fixed luminosity threshold, finding $R_{\rm sGRB}(L>10^{50}\,\mathrm{erg\,s^{-1}})=3.6^{+6.9}_{-2.5}~\mathrm{Gpc^{-3}\,yr^{-1}}$ for the full sample and $1.3^{+2.0}_{-0.7}~\mathrm{Gpc^{-3}\,yr^{-1}}$ for the flux-limited case. A subsequent analysis by \citet{pracchia_salafia2026} built upon the methodology of \citet{Salafia2023} to incorporate the impact of the delay time distribution. They derive a rate of short GRBs that is consistent with the current gravitational wave constraints \citep{gwtc4pop}, see \S \ref{sec:GW_rates} and Figure 3 of \citet{pracchia_salafia2026}.

In Figure \ref{fig:rates_lmin} we show a comparison of different available estimates of $R_{\rm sGRB}$ and their corresponding luminosities.

\subsection{Gravitational waves}
\label{sec:GW_rates}

The LVK collaboration provides independent constraints on the local merger rates of BNS and NSBH systems through GW observations \citep{gwtc4,gwtc4pop}. During the latest LVK observing run (O4), spanning about 2 years, no new high significance BNS merger has been detected. As a result, the inferred BNS merger rate has continued to shrink towards the lower end of the initial 90\% credible interval. In particular, the LVK BNS rate credible interval inferred from GWTC-4 favors substantially lower values ($22-250$ Gpc$^{-3}$ yr$^{-1}$; assuming the \textsc{FullPop-4.0} model; \citealt{gwtc4pop}) compared to earlier GWTC-3 constraints ($10-1700$ Gpc$^{-3}$ yr$^{-1}$ \citealt{KAGRA:2021duu}, derived from the \textsc{Broken Power Law + Dip} (PDB) model; \citealt{Farah_2022}). This downward shift in BNS rate estimates since GW170817 reflects the absence of additional BNS detections despite increased sensitivity and observing time. 

The LVK merger rate density we use throughout this work comes from the rate posterior calculated with the \textsc{FullPop-4.0} model \citep{gwtc4}. The \textsc{FullPop-4.0} model is based on the \textsc{Power Law + Dip} \citep{Fishbach_2020}, \textsc{Broken Power Law + Dip} \citep{Farah_2022} and \textsc{MultiPDB} \citep{Mali_2025} frameworks, and accounts for events with a primary component residing in the lower mass gap. This is particularly important for our analysis because NSBH mergers with a black hole in the lower mass gap are significantly more likely to undergo tidal disruption outside the $R_{\rm ISCO}$ \citep{2009Shibata, 2012Shibata_erratumto2009, Foucart2020,Xing_2025,Martineau_2026}. As a result, these systems are expected to dominate the population of NSBH mergers that can produce observable electromagnetic counterparts \citep{230529_everything,230529_GRB,Kunnumkai:2024qmw,Martineau_2026}.

To account for the absence of high significance BNS detections throughout O4, we rescale the BNS rate factor computed using the \textsc{FullPop-4.0} model, i.e., $R_{\rm BNS} = 89^{+159}_{-67}~\mathrm{Gpc^{-3}\,yr^{-1}}$ by a factor $B = 0.44$, yielding $R_{\rm BNS} = 39^{+70}_{-29}~\mathrm{Gpc^{-3}\,yr^{-1}}$. Errors are quoted as 90\% credible intervals. The factor $B$ is derived using the BNS ranges and duty cycles described in \url{https://gwosc.org/detector_status/}. Here $B$ is defined as:
\begin{equation}
    B = \frac{\sum{VT,_{\text{all observing runs}}}}{\sum VT, _{O1-O4a}}
\end{equation}
where $VT$ denotes the sensitive spacetime volume. This correction assumes no redshift evolution of the BNS distribution over the detectable volume, and that changes in $VT$ are constant in mass. 
While neither of these assumptions are strictly true, the BNS rate has not been found to exhibit a statistically significant redshift evolution over the observable volume by LVK, and the mass range spanned by BNSs is relatively narrow. Thus, we estimate that these assumptions add a subdominant source of error to the calculation when compared with the statistical uncertainty in the BNS rate.

The assumed NSBH merger rate density $R_{\rm NSBH} = 23^{+20}_{-13}~\mathrm{Gpc^{-3}\,yr^{-1}}$ \citep{gwtc4pop} is taken directly from LVK rate posterior and is not rescaled for non-detections.

\section{Methods}
\label{sec:methods} 

In this section, we describe our framework for connecting $R_{\rm sGRB}$ to compact binary merger populations. We consider two scenarios: (i) all sGRBs originate from BNS mergers, (ii) both BNS and NSBH mergers contribute to the sGRB population. In both the cases, we relate $R_{\rm sGRB}$ to the intrinsic compact binary merger rate $R_{\rm BNS}$ or $R_{\rm NSBH}$ through the jet launching fraction $f_{s}$ and geometric beaming corrections $\eta$.

To assess how sensitive the inferred jet opening angles and jet-launching fractions are to the observed sGRB rate density, we perform Bayesian inference with 3 different local rates from the literature. First, we adopt a local sGRB rate density of $8^{+5}_{-3}~\mathrm{Gpc^{-3}\,yr^{-1}}$, 
assuming isotropic emission, as an optimistic (i.e. on the high end of the range of rates considered) estimate of the sGRB rate \citep{Coward2012}. This rate has been previously adopted in literature \citep{Sarin2022PRD} and lies within the commonly quoted range of $\sim10~\mathrm{Gpc^{-3}\,yr^{-1}}$ \citep{GuettaPiran2006, Nakar2006, Coward2012, Fong2015ApJ815102,2023ApJ...959...13R}. However, GRB 080905A, a relatively faint but nearby event with $z \sim 0.122$, dominates the local sGRB rate density from \citet{Coward2012}. Due to the uncertainty in the redshift of this event (see Section \ref{subsec:incorrect_redshift} for a detailed discussion), excluding GRB 080905A provides a more conservative estimate, which according to \citet{Coward2012}, would give $R_{\rm sGRB} = 3^{+2}_{-1}~\mathrm{Gpc^{-3}\,yr^{-1}}$. As a representative estimate for this, we take $3^{+3.7}_{-1.9}~\mathrm{Gpc^{-3}\,yr^{-1}}$ as a fiducial case (errors are quoted as 90\% credible interval, see \citealt{1986Gehrels}). Lastly, we consider a scenario where the local sGRB rate density is $\sim 1~\mathrm{Gpc^{-3}\,yr^{-1}}$ to estimate the implications of the lowest rate densities in the literature \citep[e.g. ,][]{Ghirlanda2016,Salafia2023}.

Following \citet{Sarin2022PRD}, we relate the observed sGRB rate to BNS and NSBH merger populations within a Bayesian framework. Unlike \citet{Sarin2022PRD}, which adopted BNS and NSBH merger rate constraints from the GWTC-2 catalog \citep{LIGOScientific:2020kqk}, we adopt the most recent GW rate constraints from GWTC-4 \citep{gwtc4,gwtc4pop} and rescale the BNS rates to account for the lack of high significance BNS detection throughout the full O4 observing run. 

\subsection{Simulations}\label{sec:sims}

In addition to using the rate posterior computed from GWTC-4, we simulate a compact binary population with parameters drawn from the maximum \emph{a posteriori} population fit to GW observations in GWTC-4, using the \textsc{FullPop-4.0} population model \citep{gwtc4pop}. This population is used to provide astrophysics-informed prior for our inference. The compact binary parameters, herein dubbed ``injections'', are drawn using the public \texttt{cbc-population-distributions}\footnote{\url{https://github.com/weizmannk/cbc-population-distributions}} code \citep{Kiendrebeogo_2026}. The injections are draws from the underlying astrophysical population of GW sources, rather than the distribution of \emph{detected} GW events. The simulation set up closely follows \citet{Kunnumkai:2024tuq}; the main difference is that here we draw the binaries from GWTC-4 \textsc{FullPop-4.0} model rather than the PDB model based on GWTC-3.

We adopt the maximum \emph{a posteriori} equation of state (EoS) from \citet{2022Huth} as our fiducial EoS, for which the Tolman Oppenheimer Volkoff mass ($M_{\rm TOV}$) is 2.436 $M_{\odot}$. M$_{\rm TOV}$ is the maximum possible mass a non-rotating neutron star can take, for a given EoS. Hereafter, we refer to the lower and upper 95\% credible interval of our EoS constraints as softer and stiffer EoS respectively. See Section 2.1 of \citet{Kunnumkai:2024qmw} for details on EoS selection.

\subsection{Scenario 1: All sGRBs from BNS mergers}\label{sec:BNScase}

Under the assumption that all observed sGRBs originate from BNS mergers, the volumetric rate density of sGRBs can be expressed as
\begin{equation}
    R_{\rm sGRB} = f_{\rm s,BNS}\,\eta_{\rm BNS}\,R_{\rm BNS}
\label{eq:bns_rate}
\end{equation}
where $f_{\rm s,BNS}$ denotes the fraction of BNS mergers that successfully launch a relativistic jet, $R_{\rm BNS}$ is the local BNS merger rate density inferred from LVK GW observations, and $\eta_{\rm BNS}$ is the geometric beaming fraction, i.e., the probability that the jet is oriented toward the observer. This is given by:
\begin{equation}
f_b \equiv 1-\cos\theta_j
\end{equation}
with $\theta_{\rm BNS}$ the jet half opening angle, assuming a top hat jet.

We adopt the following priors unless otherwise stated:

{\begin{equation}
\begin{aligned}
\theta_{\rm BNS} \sim \mathcal{U}(0\degree, 30\degree) \\
R_{\rm BNS} \sim \mathrm{KDE}(R_{\rm BNS}^{\rm LVK}) \\
f_{\rm s,BNS} \sim \mathcal{U}(0.10, 0.96)
\end{aligned}
\label{eq:bns_prior}
\end{equation}}

The angular range for $\theta_{\rm BNS}$ encompasses typical jet opening angles inferred from afterglow estimates and allows for somewhat wider jets, while remaining consistent with observational constraints \citep{2023ApJ...959...13R}. The BNS rate prior is constructed from the LVK posterior samples from \textsc{FullPop-4.0} population model \citet{gwtc4pop} using a kernel density estimate (KDE), preserving the full posterior structure.

The upper bound of our fiducial choice $f_{s,\rm BNS} = 0.96$ is informed by our GW injection set described in Section \ref{sec:sims}. This bound corresponds to an optimistic scenario in which all post-merger remnants that do not result in a long-lived stable neutron star eventually collapse to a black hole and are capable of launching a jet. This includes supramassive neutron star remnants (SMNS), hypermassive neutron star remnants (HMNS), and prompt collapse systems. This is motivated by the standard sGRB models in which jet production is associated with a black hole central engine, with jet launching expected after the remnant's collapse to black hole \citep{1999Ruffert,2000Lee,2006Shibata,2006Oechslin,Murguia-Berthier_2014,Ruiz_2016,2017Ciolfi}. In such models, accretion along the spin axis of the black hole evacuates a low density funnel to support a prompt sGRB emission. 

Meanwhile, a long lived neutron star remnant remains surrounded by baryon dense medium. The high baryon pollution along the spin axis may choke the jet and suppress the Lorentz factor or prevent its formation altogether \citep{Hotokezaka_2013,Murguia-Berthier_2014,2014Nagakura,2016Siegel,2017Ciolfi,Murguia-Berthier_2017,Soares_2023}. Nevertheless, SMNS remnants are not definitively excluded as jet engines, as neutron star powered outflows can be launched in the presence of strong magnetic field and neutrino effects, and may contribute to clearing the polar region prior to collapse \citep{Dessart_2009,2015Ciolfi,Fujibayashi_2017}. Motivated by this picture, we compute the fraction of the simulated astrophysical GW population that results in a SMNS, HMNS, or prompt collapse to derive the upper limit for $f_{s,\rm BNS}$.

The lower bound of $f_{s,\rm BNS}$ reflects the expectation that at least a modest fraction of BNS mergers can produce a jet, consistent with the observed association of GW170817/GRB 170817A \citep{Abbott_2017grb,Savchenko_2017,Goldstein_2017}. 

\subsection{Scenario 2: Both BNS and NSBH as sGRB progenitors}

We now extend the framework to allow both BNS and NSBH mergers to contribute to the observed sGRB population. In this case, the local sGRB rate density is modeled as the sum of contributions from two channels:
\begin{equation}
    R_{\rm sGRB} = f_{\rm s,BNS}\,\eta_{\rm BNS}\,R_{\rm BNS}
                 + f_{\rm s,NSBH}\,\eta_{\rm NSBH}\,R_{\rm NSBH}
\label{eq:bns_nsbh_rate}
\end{equation}
Here, $R_{\rm NSBH}$ is the local NSBH merger rate, $\eta_{\rm NSBH} = 1-\cos\theta_{\rm NSBH}$ is the corresponding geometric beaming factor associated with a jet opening angle of $\theta_{\rm NSBH}$, and $f_{\rm s,NSBH}$ is the jet launching fraction for NSBH mergers. As in the BNS only case, we assume top hat jets. The priors for the BNS parameters remain unchanged, while we introduce the following additional priors for the NSBH population:
\begin{equation}
\begin{aligned}
\theta_{\rm NSBH} \sim \mathcal{U}(0\degree, 30\degree) \\
R_{\rm NSBH} \sim \mathrm{KDE}(R_{\rm NSBH}^{\rm LVK}) \\
f_{\rm s,NSBH} \sim \mathcal{U}(0, 0.23)
\end{aligned}
\label{eq:nsbh_prior}
\end{equation}

The choice of priors for $\theta_{\rm NSBH}$ and $R_{\rm NSBH}$ is similar to that of the BNS case. To place physically motivated bounds on the fraction of NSBH mergers that are capable of launching jets (namely, $f_{s,\rm NSBH}$), we estimate the fraction of mergers that retain a remnant baryon mass $M^{\rm NSBH}_{\rm rem} > 10^{-4} M_{\odot}$ outside the black hole, 10s post merger, from the GW injection set described in Section \ref{sec:sims}. We adopt this remnant mass as a conservative threshold motivated by numerical relativity studies showing that mergers capable of producing EM counterparts are typically associated with remnant mass of the order $10^{-3} - 10^{-2} M_{\odot}$ or higher \citep{Foucart2012,Hotokezaka_2013,2014Nagakura,Foucart2018,Kawaguchi_2016,Dietrich2020,Clarke_2025}. To estimate $M^{\rm NSBH}_{\rm rem}$ we use the fitting formula from \citet{Foucart2018}:
\begin{equation}
    \label{eq:mdisk_nsbh}
    \hat{M}^{\rm NSBH}_{\rm rem} = \left [{\rm max}\left( a\frac{1-2C_{\rm NS}}{\eta^{1/3}} - b~R_{\rm ISCO}\frac{C_{\rm NS}}{\eta} + c ,0\right) \right]^{1+d}.
\end{equation}
where $a=0.40642158$, $b=0.13885773$, $c=0.25512517$, $d=0.761250847$, $\eta$ = $Q/(1+Q^{2})$ refers to the symmetric mass ratio, $\hat{M}^{\rm NSBH}_{\rm rem}$ = $M^{\rm NSBH}_{\rm rem}/M^{\rm b}_{\rm NS}$, $M_{\rm NS}^{b}$ = $m_2 \left(1 + \frac{0.6C_{\rm NS}}{1-0.5C_{\rm NS}}\right)$ \citep{Lattimer_2001} and $C_{\rm NS}$ refers to the compactness of the secondary component. The upper bound of $f_{\rm s,NSBH}=0.23$ is derived assuming the stiffer EoS and the lower bound of $f_{\rm s,NSBH}=0$ reflects the possibility that all NSBH mergers fail to launch jets.

\subsection{Bayesian Inference}

We infer the parameters connecting the compact binary merger population to the observed local sGRB rate density using Bayesian inference. For each progenitor scenario, we define a parameter vector $\boldsymbol{\Theta}$. In the BNS only scenario:
\begin{equation}
    \boldsymbol{\Theta} = \{\theta_{\rm BNS}, R_{\rm BNS}, f_{s, \rm BNS}\}
\end{equation}
while in the scenario where both BNS and NSBH can be progenitors:
\begin{equation}
    \boldsymbol{\Theta} = \{\theta_{\rm BNS}, R_{\rm BNS}, \theta_{\rm NSBH}, R_{\rm NSBH}, f_{s, \rm BNS}, f_{s, \rm NSBH}\}
\end{equation}

Bayesian inference updates the prior information on $\boldsymbol{\Theta}$ using observational constraints. Bayes theorem gives 
\begin{equation}
    p(\boldsymbol{\Theta} \mid \lambda) =\frac{\mathcal{L}(\boldsymbol{\Theta} \mid \lambda)\, \pi(\boldsymbol{\Theta})}{Z}
\label{eq:bayes}    
\end{equation}
where $\pi(\boldsymbol{\Theta})$ is our prior distribution, $\mathcal{L}$ is the likelihood, and $Z$ is the evidence.

For a given parameter vector $\boldsymbol{\Theta}$, the model predicts a local sGRB rate density $R_{\rm sGRB}^{\rm model}(\boldsymbol{\Theta})$ (see Equations \ref{eq:bns_rate} and \ref{eq:bns_nsbh_rate}). We compare $R_{\rm sGRB}^{\rm model}$ to observationally motivated values of Poisson mean, $\lambda_{\rm sGRB} \in \{8,\,3,\,1\}\ {\rm Gpc^{-3}\,yr^{-1}}$, corresponding to the central values of the adopted on-axis sGRB rates $R_{\rm sGRB}$, when assuming a Poisson likelihood:
\begin{equation}
\begin{aligned}
\mathcal{L}(\boldsymbol{\Theta} \mid \lambda_{\rm sGRB})
&=
{\rm Poisson}\!\left(
k
\mid \mu = \lambda_{\rm sGRB}
\right) \\
&=
\frac{\lambda_{\rm sGRB}^{\,k}
e^{-\lambda_{\rm sGRB}}}
{k!}
\end{aligned}
\label{eq:poisson}
\end{equation}
where $k$ is the model-predicted rate density $R_{\rm sGRB}^{\rm model}(\boldsymbol{\Theta})$ and $k!$ is defined as $\Gamma(k+1)$ for a continuous $k$.

Posterior sampling is performed using nested sampling \citep{2004Skilling,Skilling2006} as implemented in \texttt{dynesty} \citep{Speagle2020}. The algorithm begins with a set of live points drawn from the prior. At each iteration, the point with the lowest likelihood is removed and replaced by a new sample with higher likelihood, causing the explored prior volume to shrink progressively toward regions favored by the data. Each discarded point is assigned a weight proportional to the product of its likelihood and change in prior volume at that step. These weighted samples collectively approximate the posterior distribution. The sum of these weights corresponds to the evidence, which is accumulated during the run. The sampling is terminated once the estimated contribution to the evidence compared to the previous step falls below a threshold value, indicating that the posterior distribution has converged.

\begin{figure}[!htpb]
    \includegraphics[scale=0.275]{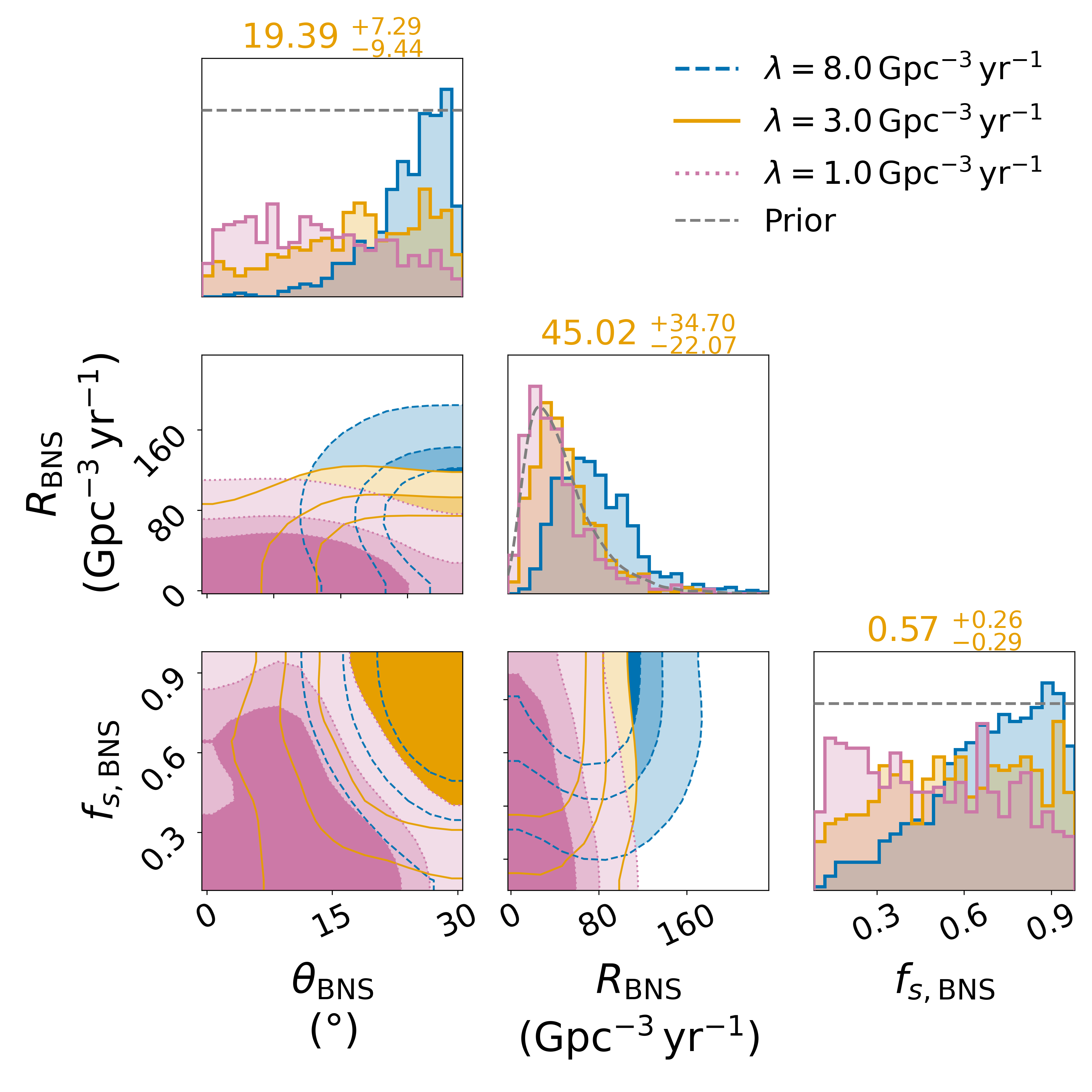}   
    \caption{Corner plot showing the posterior distributions for the binary BNS jet opening angle $\theta_{\rm BNS}$, the BNS merger rate $R_{\rm BNS}$, and the jet launching fraction $f_{\rm s,BNS}$ under the assumption that all observed sGRBs originate from BNS mergers, for different sGRB rate densities. Diagonal panels show marginalized one dimensional posteriors (histograms) along with the prior distributions (dashed lines), while off-diagonal panels show the corresponding two dimensional joint posteriors with 68\% and 90\% credible regions.}
    \label{fig:1}
\end{figure}

\begin{table*}[!htpb]
\centering
\setlength{\tabcolsep}{3.8pt}
\renewcommand{\arraystretch}{1.35}
\caption{Posterior medians and $1\sigma$ credible intervals for the BNS-only and BNS+NSBH models for different assumptions on the local short GRB rate.}
\label{tab:posterior_summary}
\begin{tabular}{c|ccc|cccccc}
\hline
\hline
 \textbf{GRB} & \multicolumn{3}{c|}{\textbf{BNS only}}
 & \multicolumn{6}{c}{\textbf{BNS + NSBH}} \\
\hline
$\lambda$
& $\theta_{\rm BNS}$
& $R_{\rm BNS}$
& $f_{s,\rm BNS}$
& $\theta_{\rm BNS}$
& $R_{\rm BNS}$
& $f_{s,\rm BNS}$
& $\theta_{\rm NSBH}$
& $R_{\rm NSBH}$
& $f_{s,\rm NSBH}$ \\

($\rm Gpc^{-3} \rm yr^{-1}$) & ($\degree$) & ($\rm Gpc^{-3} \rm yr^{-1}$) & & ($\degree$) & ($\rm Gpc^{-3} \rm yr^{-1}$) & & ($\degree$) & ($\rm Gpc^{-3} \rm yr^{-1}$) & \\
\hline

8
& $25.2^{+3.5}_{-5.5}$
& $72^{+40}_{-29}$
& $0.73^{+0.16}_{-0.25}$
& $25.1^{+3.5}_{-5.9}$
& $70^{+39}_{-29}$
& $0.73^{+0.17}_{-0.26}$
& $16^{+10}_{-11}$
& $23^{+12}_{-9}$
& $0.119^{+0.076}_{-0.081}$ \\

3
& $19.4^{+7.3}_{-9.4}$
& $45^{+35}_{-22}$
& $0.57^{+0.26}_{-0.29}$
& $18.4^{+7.8}_{-9.3}$
& $44^{+35}_{-22}$
& $0.57^{+0.27}_{-0.29}$
& $15^{+10}_{-10}$
& $23^{+11}_{-9}$
& $0.120^{+0.075}_{-0.080}$ \\

1
& $12^{+10}_{-9}$
& $36^{+32}_{-19}$
& $0.49^{+0.30}_{-0.27}$
& $12^{+10}_{-8}$
& $35^{+31}_{-19}$
& $0.49^{+0.31}_{-0.27}$
& $15^{+10}_{-10}$
& $23^{+11}_{-9}$
& $0.115^{+0.077}_{-0.078}$ \\

\hline
\hline
\end{tabular}
\end{table*}

\section{Results}
\label{sec:results}
\subsection{All sGRBs from BNS mergers}

Under the BNS-only channel hypothesis, we show the joint posteriors for $\theta_{\rm BNS}$, $R_{\rm BNS}$  $f_{s,\rm BNS}$ inferred under the three local sGRB rate density assumptions $\lambda_{\rm sGRB}\in\{8,3,1\}\,\mathrm{Gpc^{-3}\,yr^{-1}}$ in Fig.~\ref{fig:1}, with the prior assumptions in Equation \ref{eq:bns_prior}. Two features stand out:

\paragraph{(i) Wide jets for higher local sGRB rates} 
\begin{figure*}[!htpb]
    \includegraphics[scale=0.59]{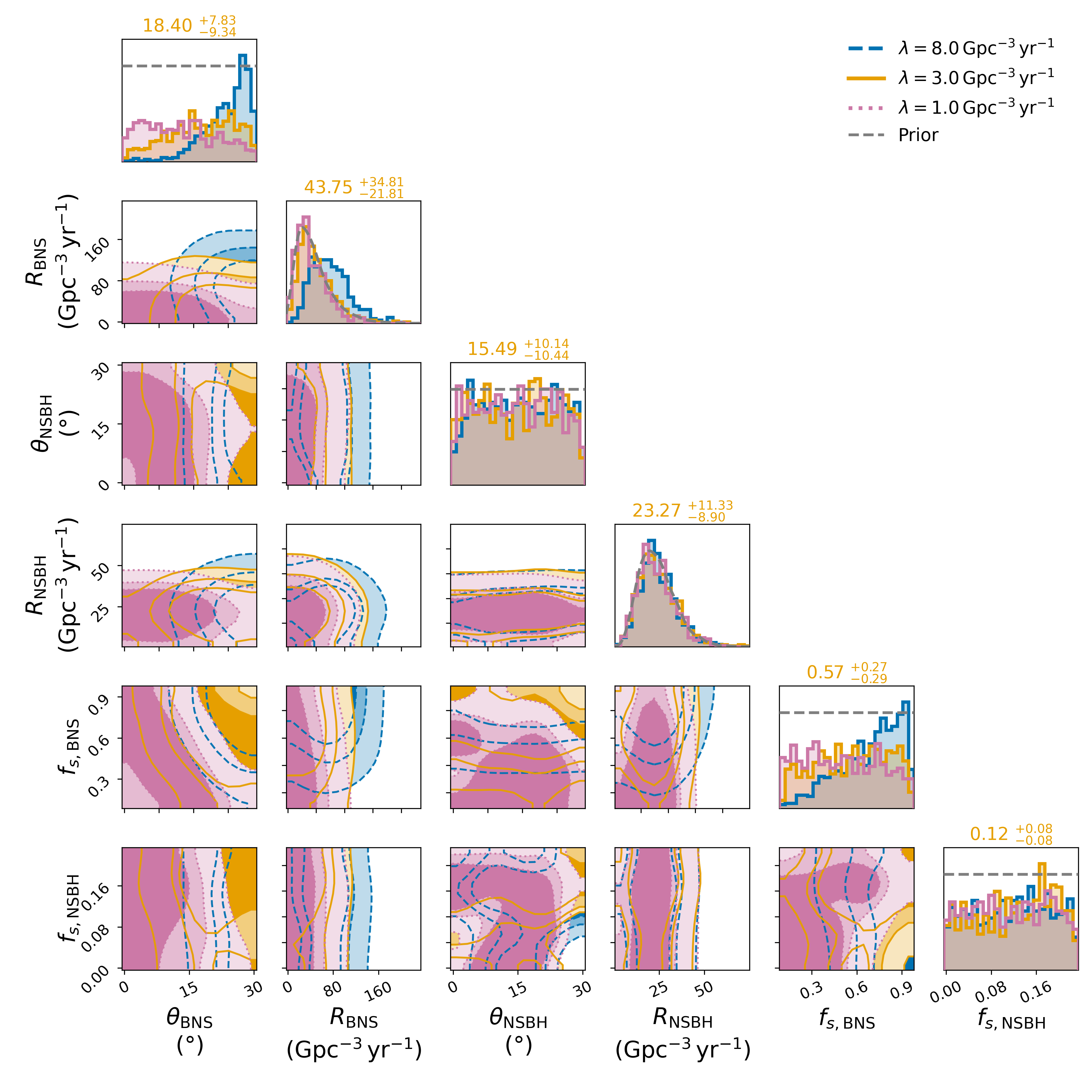}   
    \caption{Joint posterior distributions for the BNS and NSBH parameters in the two channel model. The sampled parameters include the jet opening angles $\theta_{\rm BNS}$ and $\theta_{\rm NSBH}$, the merger rates $R_{\rm BNS}$ and $R_{\rm NSBH}$, and the jet launching fractions $f_{\rm s,BNS}$ and $f_{\rm s,NSBH}$. Marginalized one dimensional posteriors are shown along the diagonal, with two dimensional correlations displayed off-diagonal. The structure of the posterior demonstrates how the addition of an NSBH channel influences the allowed parameter space while leaving the BNS parameters largely unchanged.}
    \label{fig:2}
\end{figure*}

The preferred jet half opening angle for $\lambda_{\rm sGRB} = 8~\mathrm{Gpc^{-3}\,yr^{-1}}$ \citep{Coward2012} varies from $20 - 29\degree$, (see Table \ref{tab:posterior_summary} for the exact values for all cases) much wider than canonical sGRB inferences. Instead, afterglow modeling of short GRBs typically yields opening angles of $\sim5-10\degree$ \citep[e.g.,][]{Fox2005,Grupe2006,Burrows2006,Soderberg2006,Fong2012,Fong2015ApJ815102,Berger2013jet,Troja2016jetbreak,Jin2018,Lamb2019grb160821B,Troja2019b,OConnor2021kn,Laskar2022,2023ApJ...959...13R}, with only a handful having wider $\sim15-20\degree$ jets \citep[e.g,][]{Fox2005,Grupe2006,Laskar2022,2023ApJ...959...13R}. Furthermore, the analysis of 29 sGRBs by \citet{2023ApJ...959...13R} yields a median $\theta \sim 6.1\degree$ with a substantial tail to larger values. Only 28\% of these events are consistent with $\theta \geq 10\degree$, and even fewer ($\sim 7\%$) are consistent with $\theta \geq 15\degree$. 

For $\lambda_{\rm sGRB} = 8~\mathrm{Gpc^{-3}\,yr^{-1}}$ \citep{Coward2012}, our results strongly disfavors narrow jets: $\theta_{\rm BNS} = 6\degree$ is excluded at 99.3\% credibility, while a more relaxed $\theta_{\rm BNS} = 10\degree$ is excluded at 98.5\% credibility. A median $\theta_{\rm BNS}\sim25\degree$ implies a beaming fraction
$\eta_{\rm BNS}\approx0.09$, roughly
a factor of two larger than the $\eta\!\approx\!0.04$ implied by the median $\sim16\degree$ from \citet{Fong2015ApJ815102}. Because $R_{\rm sGRB}=f_{\rm s,BNS}\eta_{\rm BNS}R_{\rm BNS}$, such wide jets require the product $f_{\rm s,BNS}R_{\rm BNS}$ to be correspondingly smaller to match a fixed observed $R_{\rm sGRB}$. Instead, our posterior simultaneously prefers large $\theta_{\rm BNS}$ and comparatively \emph{high} $f_{\rm s,BNS}$ ($\sim0.73$), indicating the sampler is exploiting the three way degeneracy among $\theta_{\rm BNS}$, $f_{\rm s,BNS}$, and $R_{\rm BNS}$ rather than settling on the narrow jet solution commonly inferred in afterglow studies. 

This tension can be futher resolved by considering lower sGRB rate densities from the literature. Excluding GRB 080905 in the calculation of rates in \citep{Coward2012}, considerably drops their inferred local sGRB rate density from $8$ to $3~\mathrm{Gpc^{-3}\,yr^{-1}}$. This allows for solutions with narrower jets, with $\theta$ ranging from 10$\degree$ to 27$\degree$, at a lower jet launching efficiency. Moreover, assuming a general scenario where the local sGRB rate density is $1~\mathrm{Gpc^{-3}\,yr^{-1}}$, to account for the lower rates predicted in the more recent literature \citep[see Figure \ref{fig:rates_lmin}; e.g.,][]{Ghirlanda2016,Salafia2023}, relives the tension between the afterglow inferred jet opening angles and those needed to explain the current LVK BNS rate estimates even further. Using $\lambda_{\rm sGRB}=1~\mathrm{Gpc^{-3}\,yr^{-1}}$ results in a 90\% credible interval jet opening angle between 4$\degree$ and 23$\degree$ with a median value of $\theta = 12 \degree$, much closer to afterglow based constraints.

We point out that results are broadly consistent with the independent inferences of \citet{desantis2026}, requiring $f_{\rm s,BNS}>0.7-0.8$ and $\theta_{\rm BNS} > 15\degree$, based on a forward-modeled population synthesis approach, which shows that there is good agreement between different, complementary approaches to addressing this issue.  

\paragraph{(ii) Merger rate}

As discussed above, observationally inferred jet opening angles are typically narrow ($\sim 5-10\degree$), implying very small beaming fractions. If all sGRBs originate from BNS mergers, such narrow jets would mean that only a tiny fraction of all BNS jets are observable from Earth, and matching the observed sGRB volumetric rate of $8~\mathrm{Gpc^{-3}\,yr^{-1}}$ would therefore require a high intrinsic BNS merger rate. For an inferred jet opening angle of $\theta = 6\degree$ \citep[e.g.][]{2023ApJ...959...13R}, assuming a maximal jet launching efficiency of $f_{s,\rm BNS} = 0.96$, reproducing $\lambda_{\rm sGRB}=8~\mathrm{Gpc^{-3}\,yr^{-1}}$ requires $R_{\rm BNS}$
$\sim 1500~\mathrm{Gpc^{-3}\,yr^{-1}}$, far above the latest GW-motivated prior range \citep{gwtc4pop}. For $\lambda_{\rm sGRB} = 1~\mathrm{Gpc^{-3}\,yr^{-1}}$ and the same assumed opening angle, Equation \ref{eq:bns_rate} would give $R_{\rm BNS} = 180~\mathrm{Gpc^{-3}\,yr^{-1}}$, consistent with GWTC-4 estimates. Using a more relaxed $\theta_{\rm BNS}$ of $10\degree$ would decrease the required BNS merger rate to reconcile the observed sGRB rate to $550~\mathrm{Gpc^{-3}\,yr^{-1}}$ for $\lambda_{\rm sGRB} = 8~\mathrm{Gpc^{-3}\,yr^{-1}}$ and $69~\mathrm{Gpc^{-3}\,yr^{-1}}$ for $\lambda_{\rm sGRB} = 1~\mathrm{Gpc^{-3}\,yr^{-1}}$, making the latter consistent with our rescaled BNS rate estimates.

The posterior results are consistent with this expectation. When $\theta_{\rm BNS}$ is inferred assuming a broad prior ($0-30\degree$), for $\lambda_{\rm sGRB}=8~\mathrm{Gpc^{-3}\,yr^{-1}}$, the posterior favors significantly wider jets ($\theta_{\rm BNS} \sim 25.2^{+3.5}_{-5.5}\,\degree$), increasing the beaming fraction and reducing the BNS merger rate to $R_{\rm BNS} = 72^{+40}_{-29}~\mathrm{Gpc^{-3}\,yr^{-1}}$ (see Figure \ref{fig:1} and Table {\ref{tab:posterior_summary}}). Nevertheless, even with these wider jets and a high jet launching efficiency of $0.73^{+0.16}_{-0.25}$, the inferred merger rate for $\lambda_{\rm sGRB}=8~\mathrm{Gpc^{-3}\,yr^{-1}}$ case is pulled towards the higher end of our rescaled BNS rates \citep{gwtc4pop}. This tension is progressively alleviated for lower sGRB rate densities, and at $\lambda_{\rm sGRB}=1~\mathrm{Gpc^{-3}\,yr^{-1}}$, the inferred $R_{\rm BNS}$ of $36^{+32}_{-19}~\mathrm{Gpc^{-3}\,yr^{-1}}$ more closely tracks the GWTC-4 prior.

\subsection{Both BNS and NSBH as sGRB progenitors}
\label{bns_nsbh_results}

\begin{figure*}[!htpb]
    \centering
    \includegraphics[scale=0.485]{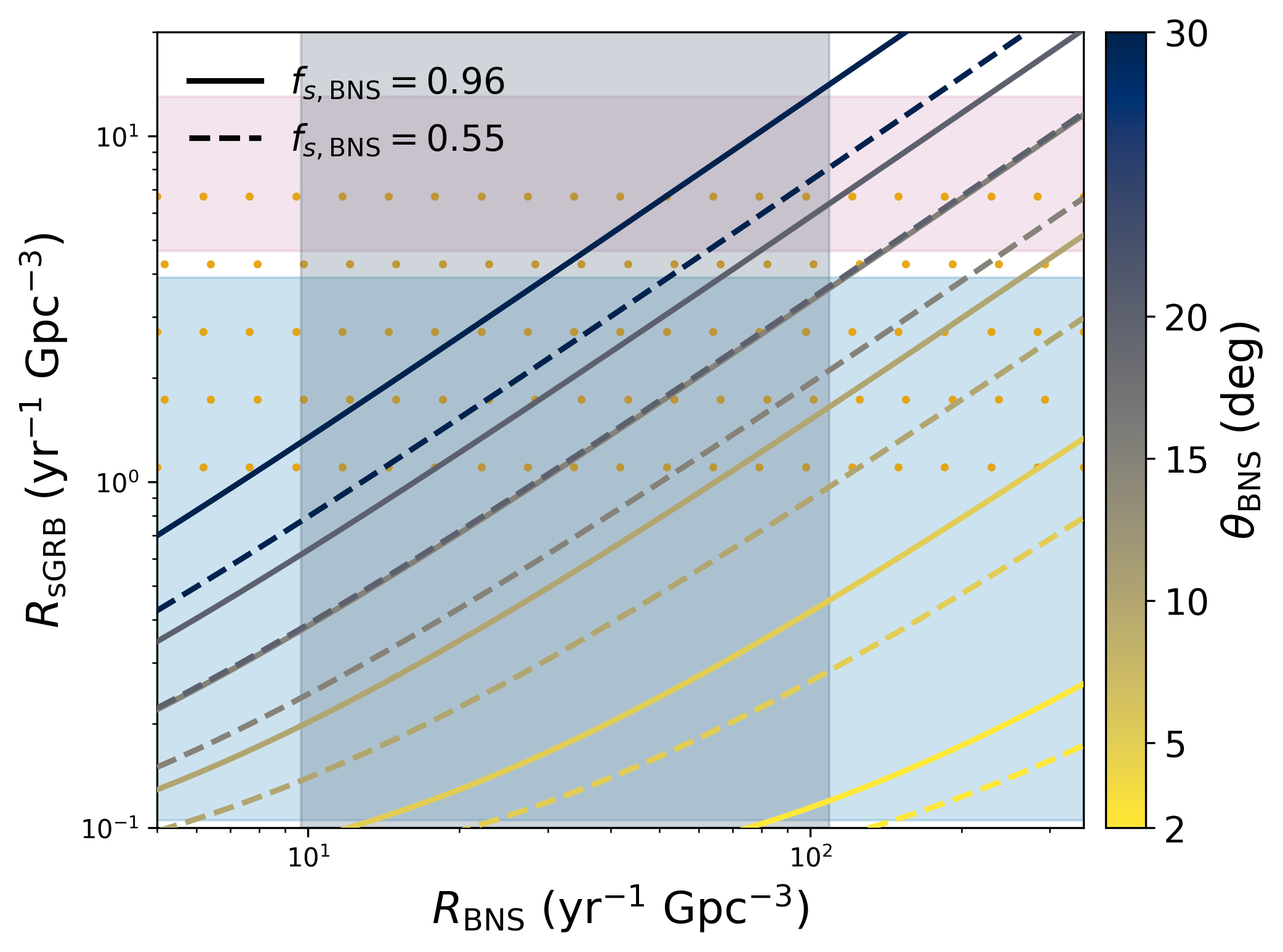}
    \includegraphics[scale=0.46]{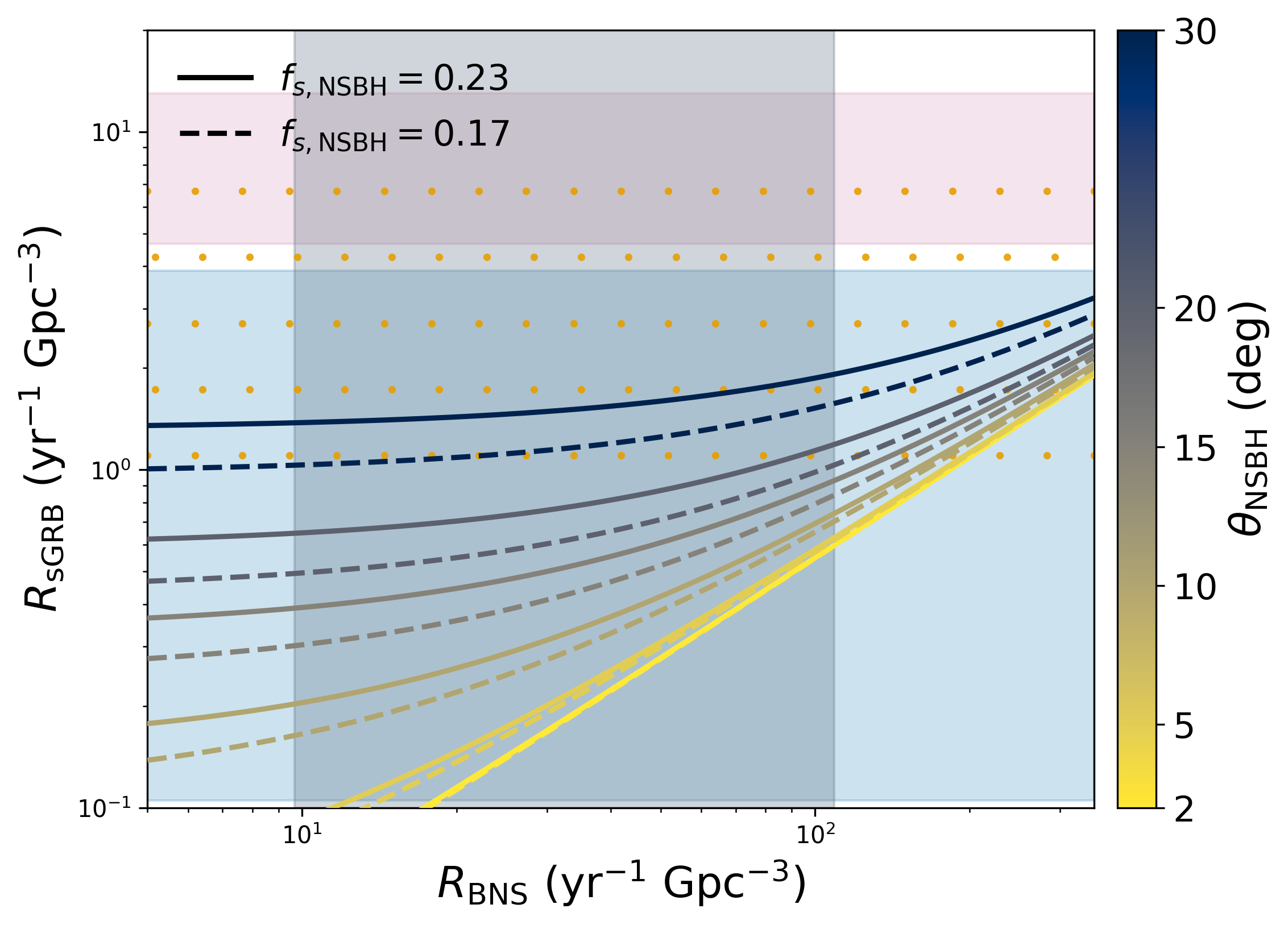}
    \caption{{$R_{\rm sGRB}$ as a function of $R_{\rm BNS}$. \textit{Left panel:} Each colored curve corresponds to different values of the BNS jet opening angle $\theta_{\rm BNS}$. The solid and dashed curves corresponds to $f_{s,\rm BNS} = 0.96$ and $f_{s,\rm BNS} = 0.55$, respectively. The NSBH contribution is fixed at $\theta_{\rm NSBH} = 6.1 \degree$ and $f_{s,\rm NSBH} = 0.23$. \textit{Right panel:} Each colored curve correspond to a different NSBH jet opening angle $\theta_{\rm NSBH}$, with solid and dashed lines corresponding to $f_{s,\rm NSBH} =0.23$ and $f_{s,\rm NSBH}=0.17$ respectively. The BNS parameters are fixed at $\theta_{\rm BNS} = 6.1 \degree$ and $f_{s,\rm BNS} = 0.96$.
    Horizontal regions represents the 90\% credible intervals of the local sGRB rate density estimates adopted in this work, with the dotted region corresponding to the  case of $\lambda_{\rm sGRB} = 3~\mathrm{Gpc^{-3}\,yr^{-1}}$. Vertical shaded regions show the LVK inferred BNS merger rate constraints rescaled to account for no BNS detections in O4. The intersections between the model curves and the shaded bands identify the combinations of jet opening angles and successful jet launching fractions that are consistent with both GW merger rates and the observed sGRB rate density.}}
    \label{fig:rgrb_rbns_thetabns_lines}
\end{figure*}

While sGRBs are mainly associated with BNS mergers, theoretical studies demonstrate that NSBH mergers can launch relativistic jets under favorable conditions, and thus present a viable additional progenitor channel \citep{Mochkovitch1993Nature,VossTauris2003MNRAS,Rosswog_2005, Hotokezaka_2013, Paschalidis_2015,Ciolfi2018sGRBReview}. Observationally, there is no conclusive evidence for a distinct NSBH population, but certain subclasses such as extended emission sGRBs exhibit properties \citep{Norris2006}, such as their longer duration, that have been suggested as tentative signatures of an additional progenitor channel \citep{2008Troja,Gompertz_2020,Zhu_2022}. Although claims of extended sGRBs coming from a separate channel from regular sGRBs have been challenged \citep{2013FongBerger, Li_2025}, given the theoretical plausibility, we extend the rate model to Equation \ref{eq:bns_nsbh_rate}. Figure \ref{fig:2} shows the joint posteriors for $\theta_{\rm BNS},R_{\rm BNS},\theta_{\rm NSBH},R_{\rm NSBH},f_{\rm s,BNS},f_{\rm s,NSBH}$. A few features stand out:

\paragraph{(i)} Including the NSBH channel does not significantly affect the BNS parameters relative to the BNS-only model. For $\lambda_{\rm sGRB} = 8~\mathrm{Gpc^{-3}\,yr^{-1}}$, 
$\theta_{\rm BNS}$ remains near $\sim 19-29\degree$, $R_{\rm BNS}$ stays around $70^{+39}_{-29}~\mathrm{Gpc^{-3}\,yr^{-1}}$ (pulled towards the higher end of the GWTC-4 prior) and $f_{\rm s,BNS}$ is broad with median near $\sim 0.73$, identical to the BNS-only scenario. A behavior similar to the BNS-only model is seen for lower sGRB rates as well, as evident in Table \ref{tab:posterior_summary}. This indicates that the addition of the NSBH channel does not relieve the requirement that BNS mergers account for the majority of the observed sGRB progenitors. This is expected given the NSBH lower volumetric rate and likely lower jet launching efficiency.

\paragraph{(ii)} The NSBH parameters are only weakly constrained, if at all, and show distributions that are very close to the priors. Notably, the NSBH rate closely matches the LVK GWTC-4 prior across all $\lambda_{\rm sGRB}$ values, and $f_{s,\rm NSBH}$ remains broadly distributed within the prior range. As for point (i), this is not unexpected given that the NSBH volumetric rate is a factor of a few lower relative to the BNS one, and it can only partially contribute to the sGRB rate because the majority of NSBHs do not disrupt the neutron star outside of the remnant ISCO, so the NSBH population does not contribute to the majority of the sGRBs and cannot be meaningfully constrained with the available data. 

To quantify the relative contribution of the BNS and NSBH channels to the observed on-axis sGRB rate, we rewrite Equation \ref{eq:bns_nsbh_rate} as $R_{\rm sGRB} = R^{\rm BNS}_{\rm sGRB} + R^{\rm NSBH}_{\rm sGRB}$. For each posterior sample, we then compute the individual BNS and NSBH contributions to $R_{\rm sGRB}$. Assuming fiducial priors of $f_{s,\rm BNS}\sim U(0.10,0.96)$ and $f_{s,\rm NSBH}\sim U(0,0.23)$, we find that the observed sGRB population is strongly dominated by the BNS channel across the full range of $\lambda_{\rm sGRB}$ values considered. The median NSBH contribution increases from $1.6^{+6.8}_{-1.5}\%$ for $\lambda_{\rm sGRB} = 8~\mathrm{Gpc^{-3}\,yr^{-1}}$, to $6.3^{+31.3}_{-5.9}\%$ for $\lambda_{\rm sGRB} = 3~\mathrm{Gpc^{-3}\,yr^{-1}}$, to $16^{+57}_{-15}\%$ for $\lambda_{\rm sGRB} = 1~\mathrm{Gpc^{-3}\,yr^{-1}}$. Correspondingly, the BNS channel contributes $98.4^{+1.5}_{-6.9}\%$, $93.7^{+5.9}_{-31.3}\%$, $84^{+15}_{-57}\%$ for $\lambda_{\rm sGRB} \in \{8,\,3,\,1\}\ {\rm Gpc^{-3}\,yr^{-1}}$ respectively. The values quoted above are medians with $1\sigma$ bounds.

Next, we explore how the different BNS and NSBH parameters may explain any possible discrepancy between the sGRB and GW rates. Figure \ref{fig:rgrb_rbns_thetabns_lines} shows the predicted local sGRB rate density as a function of intrinsic BNS merger rate (left panel). The lines correspond to different BNS jet opening angles, where the solid line corresponds to a jet launching fraction of 0.96 (maximum number of injected BNS mergers in our simulation that would result in a SMNS, HMNS or a prompt collapse black hole remnant) and the dashed line corresponds to 0.55. We choose $f_{s,\rm BNS} = 0.55$ to consider a restrictive scenario in which only the prompt collapse and HMNS remnants are assumed to launch jets. This is motivated by numerical simulations showing that long lived remnants can drive sustained baryon loaded winds along the polar region, which suppresses the jet Lorentz factor and can prevent successful jet breakout on sGRB timescales \citep{1999Ruffert,2002Rosswog,Dessart_2009,Murguia-Berthier_2014,Murguia-Berthier_2017, 2017Ciolfi}. We fix $R_{\rm NSBH} = 43~\mathrm{Gpc^{-3}\,yr^{-1}}$, which is the upper end of GWTC-4 NSBH rate estimate with \textsc{FullPop-4.0} model. The horizontal regions correspond to different $\lambda_{\rm sGRB}$ values used in this study and their 90\% credible intervals from \citet{1986Gehrels}. The rose shaded region corresponds to $1^{+3}_{-1}~\mathrm{Gpc^{-3}\,yr^{-1}}$, the blue shaded region corresponds to $8^{+5}_{-3}~\mathrm{Gpc^{-3}\,yr^{-1}}$ and the dotted region in the middle corresponds to $3^{+4}_{-2}~\mathrm{Gpc^{-3}\,yr^{-1}}$. 

It is evident from Figure \ref{fig:rgrb_rbns_thetabns_lines} that the higher sGRB rates corresponding to $\lambda_{\rm sGRB}=8~\mathrm{Gpc^{-3}\,yr^{-1}}$ require a wide BNS jet opening angle ($\theta_{\rm BNS} > 17\degree$) to match the current LVK BNS rate constraints (rescaled to account for no BNS detection in O4; vertical shaded region), even if we consider 96\% of BNS mergers to result in a sGRB. This requirement becomes relaxed when we consider $\lambda_{\rm sGRB}=3~\mathrm{Gpc^{-3}\,yr^{-1}}$, as $\theta_{\rm BNS} > 7\degree$ ($\theta_{\rm BNS}\geq 11\degree$) could explain the BNS (rescaled) rates if at least 55\% of 
BNS successfully launch a jet. The blue shaded region corresponding to $\lambda_{\rm sGRB}=1~\mathrm{Gpc^{-3}\,yr^{-1}}$ is consistent with the GW BNS rate constraints down to $\theta_{\rm BNS}\geq3\degree$.

For what concerns the NSBH parameters (right panel), assuming fiducial parameter values of $\theta_{\rm BNS} = 6.1 \degree$ \citep{2023ApJ...959...13R}, $f_{s,\rm BNS} = 0.96$ and $R_{\rm NSBH} = 43~\mathrm{Gpc^{-3}\,yr^{-1}}$, we explore the allowed NSBH contribution for two representative choices of jet launching fractions, $f_{s,{\rm NSBH}} = 0.23$ (solid lines) and $f_{s,{\rm NSBH}} = 0.17$ (dashed lines). These values incorporate the uncertainty in tidal disruption and in the amount of disk mass remaining outside the black hole after merger. Within this range, even with the largest NSBH jet angles we consider ($\theta_{\rm NSBH}$ = $30\degree$), we find that the higher sGRB rates (e.g., $\lambda_{\rm sGRB} = 8~\mathrm{Gpc^{-3}\,yr^{-1}}$) cannot be reconciled with the observed GWTC-4 BNS rates \citep{gwtc4pop}. On the other hand, for $\lambda_{\rm sGRB} = 1~\mathrm{Gpc^{-3}\,yr^{-1}}$ the current GW BNS rates can be reconciled even with the lowest jet opening angles ($\theta_{\rm NSBH}$ = $2\degree$). Our inferred range of $f_{s,\rm NSBH}$ is consistent with recent population studies; \citet{Clarke_2025} finds that $\sim 19\%$ of NSBH mergers in their simulated population with remnant mass $>0$ undergoes tidal disruption.

\section{Discussion}
\label{sec:discussion}

In what follows, we discuss various sources of systematic uncertainties that may have an impact on the sGRB rates and, as a consequence, on our conclusions.

\subsection{Dependence of the results on the assumed GW and sGRB rates}

Our BNS merger rate prior is based on the GWTC-4 population analysis \citep{gwtc4pop}, rescaled to account for the continued non-detection of high significance BNS mergers through the end of O4, as discussed in Section \ref{sec:GW_rates}. We note that the detection of a sub-threshold BNS merger candidate GW231109\_235456 \citep{niu2025gw231109,2025Wouters}, yields an inferred BNS merger rate of $53 - 342~\mathrm{Gpc^{-3}\,yr^{-1}}$. While this estimate depends on the true astrophysical nature of the candidate, it could mean that the true BNS rate is a factor of few higher than the rescaled BNS rate considered in this study. Such BNS rates would still require fairly wide jets ($\theta_{\rm BNS} > 10\degree$ for $f_{s,\rm BNS} = 0.96$) to explain high local sGRB rates (e.g., $\lambda_{\rm sGRB} = 8~\mathrm{Gpc^{-3}\,yr^{-1}}$). 

In addition, we note the possible existence of compact object merger candidates with sub-solar mass components \citep{GCN250818kLVK,2025GCNS251112cm}, one of which also had a candidate EM counterpart \citep{Kasliwal_2025,  hall2025,hall2025desi}. If BNS mergers involving component masses $< 1~M_{\odot}$ exist, \citet{Fishbach_2026} reports an upper limit for the rate of such objects that is $\sim 2-3$ times their BNS rate constraint of $28-300~\mathrm{Gpc^{-3}\,yr^{-1}}$, inferred from their analysis on GW data from \citet{gwtc4}. This is consistent with the fact that current subsolar mass merger rate estimates are still too broad and unconstrained \citep{kacanja2026subsolar} to provide meaningful constraints in our analysis, but future analyses of the aforementioned candidates will provide further insight on these populations and their rates. In this work, we do not consider these objects as BNS mergers, although if they contribute to the sGRB population they could potentially be part of the non-standard progenitors (see Section \ref{non-std prog}).

\begin{figure*}[!htpb]
    \includegraphics[scale=0.39]{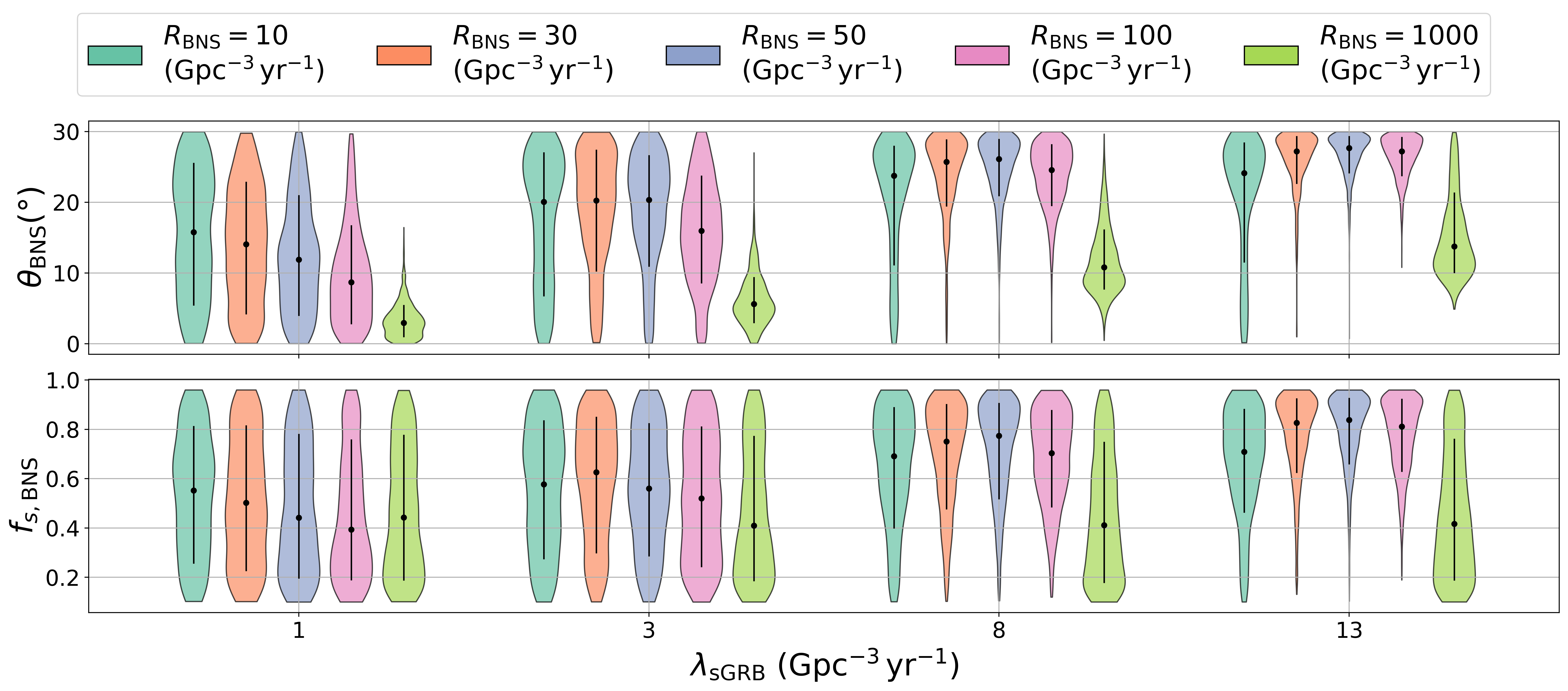}  
 
    \caption{Violin plots showing the inferred posterior distributions of the BNS jet opening angle $\theta_{\rm BNS}$ and jet launching fraction $f_{\rm s,BNS}$ for different assumed sGRB Poisson rates $\lambda_{\rm sGRB}$ and a range of intrinsic BNS merger rates $R_{\rm BNS}$, corresponding to the BNS only scenario. The variation across colors illustrates how changes in the assumed $R_{\rm BNS}$ influence the allowed $\theta_{\rm BNS}$–$f_{\rm s,BNS}$ parameter space for the BNS-only progenitor hypothesis. The black dot marks the posterior median, while the vertical black lines indicate 16th-84th percentile credible interval. }
    \label{fig:vp_Rbns}
\end{figure*}

It is also interesting to compare the current BNS rates with the estimates from previous GW releases, such as $170^{+270}_{-120}$ Gpc$^{-3}$ yr$^{-1}$ from GWTC-3 \citep{KAGRA:2021duu}, $320^{+490}_{-240}$ Gpc$^{-3}$ yr$^{-1}$ from GWTC-2 \citep{LIGOScientific:2020kqk}, and even the estimate following the detection of GW190425 ($250-2810$ Gpc$^{-3}$ yr$^{-1}$) \citep{LIGOScientific:2020aai}. The rate constraints have been shrinking over time towards lower values, from O($10^{3}$) Gpc$^{-3}$ yr$^{-1}$, consistent with the initial GW rates estimates, to 10 Gpc$^{-3}$ yr$^{-1}$ (lower limit of rescaled BNS rate estimates), so we explore how such evolution impacts our conclusions. 

Figure~\ref{fig:vp_Rbns} illustrates how the inferred BNS jet opening angle and jet launching fraction respond to different combinations of the intrinsic BNS merger rate $R_{\rm BNS}$ and the assumed local sGRB rate density, represented by a Poisson distribution with mean $\lambda_{\rm sGRB}$, in units of $\mathrm{Gpc^{-3}\,yr^{-1}}$. For very high intrinsic merger rates ($R_{\rm BNS}\sim 1000~\mathrm{Gpc^{-3}\,yr^{-1}}$), the inferred jet opening angle is highly sensitive to the observed sGRB rate. If $\lambda_{\rm sGRB}=1~\mathrm{Gpc^{-3}\,yr^{-1}}$, the posterior pushes $\theta_{\rm BNS}$ towards small values ($\lesssim5\degree$). In this regime, the large number of mergers combined with a low observed sGRB rate density implies that we are effectively only detecting on axis events with sGRBs, so the model compensates by requiring extremely narrow jets. In contrast, at the same high intrinsic merger rate but with a larger assumed sGRB rate density (e.g., $\lambda_{\rm sGRB}\sim13~\mathrm{Gpc^{-3}\,yr^{-1}}$), the inferred opening angle increases to $\theta_{\rm BNS}\sim15\degree$, indicating that off-axis sGRBs must also be contributing to the observed sample.

At more moderate BNS rates closer to current LVK constraints ($R_{\rm BNS}\sim 10-50~\mathrm{Gpc^{-3}\,yr^{-1}}$), the violin plots show that if we detect only sGRBs following $\lambda_{\rm sGRB}=1 ~\mathrm{Gpc^{-3}\,yr^{-1}}$, the jet angle remains consistent with relatively narrow ($\theta_{\rm BNS}\lesssim 15\degree$) jets. In contrast, for $\lambda_{\rm sGRB}\sim8~\mathrm{Gpc^{-3}\,yr^{-1}}$, the posterior shifts toward $\theta_{\rm BNS}\sim 30\degree$ with $f_{\rm s,BNS}\sim 0.6-0.8$. In this case, reproducing the observed sGRB rate density with neutron star mergers alone requires both a substantial jet-launching efficiency and jet opening angles wider than typically inferred from afterglow modeling. 

Comparing the BNS-only and BNS+NSBH scenarios, we find that the inferred BNS jet properties are largely insensitive to the inclusion of the NSBH channel across the range of $(R_{\rm BNS},\lambda_{\rm sGRB})$ explored here (see also \ref{bns_nsbh_results}). In both cases, the dominant driver of the inferred opening angle is the assumed local sGRB rate density $\lambda_{\rm sGRB}$: lower values remain consistent with relatively narrow jets ($10-20\degree$), while higher values push the posterior towards wider jets ($\theta_{\rm BNS}\geq 25\degree$). While increasing $f_{s,\rm NSBH}$ could in principle enhance the NSBH contribution, such high jet-launching efficiencies are not supported by our simulated NSBH merger population.

\subsection{Additional possible progenitor channels}
\label{non-std prog}

An additional source of uncertainty is the possible contribution from channels beyond BNS and NSBH mergers to the observed sGRB rate. We consider this possibility as a consistency check by determining the required properties of other channels to significantly affect our conclusions. Throughout this work the adopted $R_{\rm sGRB}$ values correspond to the classical non-collapsar sGRB population. The adopted sGRB samples account for the fact that a simple $T_{90}$ based classification is detector-dependent and can lead to significant collapsar contamination, particularly for \emph{Swift} bursts \citep{2013Bromberg}. Indeed, the samples considered in Section \ref{sec:GRB_rates} use probabilistic classifications that account for burst duration, spectral properties, and detector bandpass to estimate whether an event is a non-collapsar or short-duration collapsar \citep{Coward2012,Wanderman2015}. As a consequence, we do not believe that the GRB rates estimates we assume should be significantly contaminated by collapsars.

While it is believed that compact mergers may also produce long duration or extended emission events (e.g., GRB 211211A, \citealt{Rastinejad2022Nature,2022Natur.612..228T,2022Natur.612..232Y}, and GRB 230307A, \citealt{Levan2023,Yang2024,Gillanders2023,Gillanders2025}), such events are not included in the rate estimates used in this work.

Previous studies have associated white dwarf-black hole (WDBH) and white dwarf-neutron star (WDNS) mergers with long duration GRBs \citep{1999FryerWDBH,2024Chen,2024Lloyd,2025Cheong,2025Liu,Chrimes2025,2025Chen}, rather than the classical short duration GRB population considered here, making them an unlikely contributor to the observed sGRB rate. Nevertheless, we quantify what properties such channels (or similarly, any other additional channel beyond BNS and NSBH mergers)  would require in order to contribute significantly to the non-collapsar sGRB rates and explain the results obtained for the higher sGRB rates considered ($\lambda_{\rm sGRB} =  8~\mathrm{Gpc^{-3}\,yr^{-1}}$).

To quantify what it takes for a WDBH or non-standard channel to contribute significantly, we can use the same rate model employed throughout the paper,
\begin{equation}
R_{\rm sGRB}=\sum_i f_{s,i}\,\eta_i\,R_i,\qquad \eta_i\equiv 1-\cos\theta_i
\end{equation}
where the sum covers all channels $i$, and for simplicity, we assume $f_{s,\rm BNS}=0.96$, $f_{s,\rm NSBH}=0.23$, $\theta_{\rm BNS}=\theta_{\rm NSBH}=6.1\degree$, $R_{\rm BNS}=39~\mathrm{Gpc^{-3}\,yr^{-1}}$ (rescaled median BNS rate accounting for no high-significance BNS detection in O4), $R_{\rm NSBH}=23~\mathrm{Gpc^{-3}\,yr^{-1}}$ (median NSBH rate from \textsc{FullPop-4.0} model;  \citealt{gwtc4pop}). This yields a combined $\rm BNS + \rm NSBH$ contribution $R_{\rm sGRB}^{\rm (BNS+NSBH)}
\simeq 0.242~\mathrm{Gpc^{-3}\,yr^{-1}}$. For a WDBH rate of $R_{\rm WDBH}=6~\mathrm{Gpc^{-3}\,yr^{-1}}$ \citep{Chrimes2025}, even matching an sGRB rate density of $1~\mathrm{Gpc^{-3}\,yr^{-1}}$ assuming $\theta_{\rm WDBH}=6\degree$ would require jet launching fractions far exceeding unity, which is unphysical. Even if we assume maximal jet launching efficiency ($f_{s,\rm WDBH}=1$), the required jet opening angles remain extremely large ($\sim29\degree$ for $R_{\rm sGRB} = 1~\mathrm{Gpc^{-3}\,yr^{-1}}$), $\sim57\degree$ for $R_{\rm sGRB} = 3~\mathrm{Gpc^{-3}\,yr^{-1}}$), while $R_{\rm sGRB} = 8~\mathrm{Gpc^{-3}\,yr^{-1}}$ cannot be explained with such a WDBH contribution. We conclude that WDBHs, or any other channel which may occur at a similar rate, cannot significantly contribute to the sGRB population without invoking unrealistically wide jets.

Repeating the same exercise for WDNS channel with $R_{\rm WDNS}=100~\mathrm{Gpc^{-3}\,yr^{-1}}$, see Figure 6 of \citet{Chrimes2025}, leads to less extreme but still restrictive requirements. For narrow jets $\theta_{\rm WDNS}=6\degree$, the required jet launching fraction again exceed unity. Imposing $f_{s,\rm WDNS}=1$ would imply a jet opening angle of $\sim7^\circ$, $\sim14^\circ$, and $\sim23^\circ$ for sGRB rates $1$, $3$ and $8~\mathrm{Gpc^{-3}\,yr^{-1}}$ respectively. 

Although these values are less extreme than the WDBH case due to larger assumed WDNS volumetric rate, they still require moderate to wide jets to make meaningful contribution to sGRB rates. Consequently, for higher $\lambda_{\rm sGRB}$ values, introducing such potential additional progenitors does not recover narrow jet opening angles. For comparison, afterglow based estimates typically find $\theta\sim5-10\degree$ \citep[e.g.,][]{Fox2005,Grupe2006,Burrows2006,Soderberg2006,Fong2012,Fong2015ApJ815102,Berger2013jet,Troja2016jetbreak,Jin2018,Lamb2019grb160821B,Troja2019b,OConnor2021kn,Laskar2022,2023ApJ...959...13R}. To summarize, an additional channel alone cannot reconcile some of the highest sGRB rates found ($R_{\rm sGRB} > 3~\mathrm{Gpc^{-3}\,yr^{-1}}$). So the lower bound of the observed sGRB rates remain a more feasible option to explain the GW-GRB connection.

\subsection{Short GRB jet opening angles: A selection bias against wide jets?}

\begin{figure*}
    \includegraphics[scale=0.78]{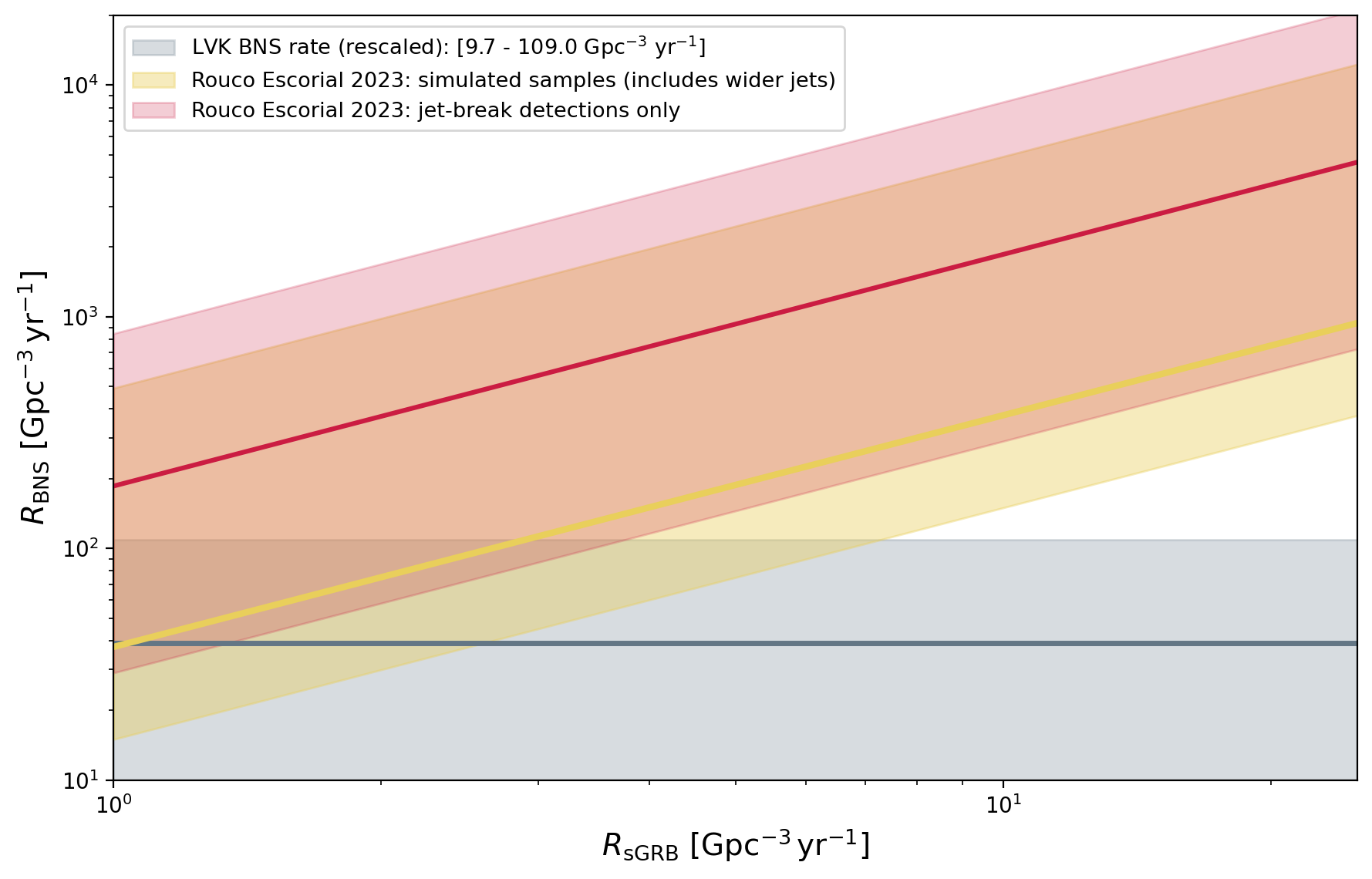}
    \caption{BNS merger rate required to reproduce the observed sGRB rate. The solid orange line shows the median intrinsic merger rate inferred from the mock jet population of \citet{2023ApJ...959...13R}, while the solid pink line shows the median rate inferred using only bursts with measured jet breaks, which preferentially selects narrow jets. The shaded regions indicate the corresponding uncertainties. The grey band shows the LVK BNS merger rate constraints rescaled to account for no BNS detection in O4. 
}
\label{fig:rescaled_rates}
\end{figure*}

Jet opening angles are inferred from temporal breaks in afterglow light curves, which occur once the relativistic outflow slows down enough that the edge of the jet becomes visible \citep{1999Rhoads,1999Sari}. In practice, such measurements are available for only a small fraction of sGRBs with well-sampled afterglows, while many events provide only lower limits due to the absence of detected jet breaks \citep[e.g.][]{Fong2015ApJ815102,2023ApJ...959...13R}. The inferred distribution spans from a few degrees for bursts with measured jet breaks to lower limits of $\geq 5-25\degree$ for events with no jet break detections \citep{Fong2015ApJ815102,2023ApJ...959...13R}. The probability that a jet is oriented towards the observer scales with the solid angle of the jet. Wider jets therefore occupy a larger fraction of the sky and should be more commonly observed as prompt GRBs. However, their jet breaks occur later and at lower fluxes and are harder to detect \citep{1999Sari,1999Rhoads}. This could mean that wide jets are systematically underrepresented in samples with measured jet opening angles and could bias our understanding of jet launching fractions and rates.

In the case of \citet{2023ApJ...959...13R}, the analysis distinguishes between \textit{i}) a sample restricted to bursts with measured jet breaks, which is biased towards narrow jets and \textit{ii}) a mock sample constructed to account for jets that lack jet break detections. In the second case, they use a ``mock'' Monte Carlo population modeling to infer the intrinsic merger rate required to reproduce the observed jet opening angle distribution. In this mock sample, they combine bursts with measured opening angles and additional wide jets ($\theta_{j} > 10\degree$, representing events with only lower limits on $\theta_{j}$), centered around $\theta_{j}\sim20\degree$.  The authors adopt a fiducial observed sGRB rate of $10~\mathrm{Gpc^{-3}\,yr^{-1}}$ following \citet{Nakar2006,Fong2015ApJ815102} and show that the inferred intrinsic merger rate scales linearly with this choice, yielding $R_{\rm mock} = 361^{+4367}_{-217}~\mathrm{Gpc^{-3}\,yr^{-1}}$, consistent with LVK BNS rates at the time. 

We show the impact of the observational bias towards wider jet opening angle in our analysis in Figure \ref{fig:rescaled_rates}, where we rescale the intrinsic rates inferred from \citet{2023ApJ...959...13R} to different observed sGRB rates, $R_{\rm sGRB}$. When only bursts with measured jet breaks are considered, the preference for narrow jets leads to large beaming corrections and would therefore require BNS merger rates $R_{\rm BNS}$ that exceeds current GW BNS constraints for most of the parameter space. In contrast, the mock sample, which includes wider jets inferred for events lacking jet break detections, yields substantially lower intrinsic rates and remains compatible with GW measurements across a wider range of parameter space. Our results are consistent with both \citet{Fishbach_2026} and \citet{desantis2026}. These authors also find that wider jets or lower $R_{\rm sGRB}$ would be needed to reconcile the current BNS and sGRB rate constraints. The overlap region between the mock sample and the rescaled LVK BNS rate in Figure \ref{fig:rescaled_rates} suggests consistency for $R_{\rm sGRB} < 7~\mathrm{Gpc^{-3}\,yr^{-1}}$ whereas the sample with only jet-break detections would require $R_{\rm sGRB} < 4~\rm Gpc^{-3}\,yr^{-1}$ to match the rescaled LVK BNS rates. These results suggest that the apparent tension between sGRB rates and GW inferred merger rates largely disappears when accounting for the population of wide jets, supporting an observed sGRB rate density of up to $7~\mathrm{Gpc^{-3}\,yr^{-1}}$. Further work should be done to properly account for the selection bias against measuring wide jets in the observed population of sGRB afterglows. 

\subsection{Impact of incorrect redshifts}
\label{subsec:incorrect_redshift}

The inferred observed volumetric rate of sGRBs from flux-limited samples depends on the assumed redshifts of individual bursts. In the $V_{\max}$ framework adopted by \citet{Coward2012}, each burst contributes to the local rate density according to the maximum comoving volume within which it could have been detected. The maximum detectable luminosity distance $d_{L,\max}$ depends on the observed peak flux, the luminosity distance of the burst, and the $k$-correction. Consequently, the inferred isotropic-equivalent luminosity of a burst, and therefore the volume $V_{\max}$ over which it would remain detectable, are both functions of the assumed redshift. If a burst is placed at a larger redshift than initially assumed, its inferred luminosity increases, which generally increases $V_{\max}$ and reduces its contribution to the inferred local volumetric rate. 

Most sGRB redshifts are assigned through host-galaxy association \citep[e.g,][]{2010Berger,2013Fong,Fong2013other,Tunnicliffe2014,OConnor2022,2022Fong}. Candidate hosts are ranked by their probability of chance coincidence, $P_{cc}$ \citep{2002Bloom}. Since $P_{cc}$ depends on the angular separation between the burst and the galaxy, as well as on the apparent brightness of the galaxy, the host with the minimum $P_{cc}$ is taken as the most probable association \citep{2002Bloom,2010Berger}. However, the $P_{cc}$ method naturally favors apparently brighter, low-$z$ galaxies. 

A particularly striking example is GRB~080905A \citep{2008GCN..8185....1P}. In the \emph{Swift} sample analyzed by \citet{Coward2012}, this burst contributes $4.9~\mathrm{Gpc^{-3}\,yr^{-1}}$ to the inferred isotropic local rate density, whereas most other events contribute between $\sim 0.04$ and $2~\mathrm{Gpc^{-3}\,yr^{-1}}$. This large contribution arises because GRB~080905A is associated with a nearby host galaxy at $z=0.122$, making it the lowest-redshift event in their sample. As a result, its accessible detection volume $V_{\max}$ is small, leading to a large weight in the $V_{\max}^{-1}$ rate estimator. Indeed, \citet{Coward2012} demonstrate that excluding this single event reduces their inferred isotropic rate from $8^{+5}_{-3}$ to $3^{+2}_{-1}~\mathrm{Gpc^{-3}\,yr^{-1}}$, illustrating how strongly the inferred rate can depend on individual nearby bursts.

The host association, and therefore the redshift, of GRB~080905A is not entirely secure. As noted by \citet{Coward2012} in their Table~1 (see note ``a''), the burst is a strong outlier to the $E_{\rm p}$--$L_{\rm p}$ \citep{Yonetoku} relation if the redshift $z=0.122$ is adopted. \citet{gruber2012} similarly identify GRB~080905A as an outlier under this low-redshift assumption and show that placing the burst at a higher redshift, $z \gtrsim 0.8$, would restore consistency with the Yonetoku relation. If the burst were instead associated with a more distant galaxy, the inferred luminosity would be significantly larger and the maximum detectable volume correspondingly greater. In this case, the contribution of GRB~080905A to the local volumetric rate would be dramatically reduced, rendering its contribution to the total rate essentially negligible. Additionally, \citet{Howell_coward2013} find a $\sim 1\%$ probability for GRB~080905A to have occurred at $z=0.122$.

This example illustrates that a single nearby burst can have a disproportionate impact on empirical rate estimates derived from small samples. A case-by-case inspection of the other bursts in the \citet{Coward2012} sample suggests that the remaining host galaxy associations are generally robust. Consequently, GRB~080905A remains the dominant source of systematic uncertainty in the \citet{Coward2012} estimate. Because this event is included in essentially all post-2008 empirical sGRB rate studies (e.g., \citealt{Wanderman2015}), any potential misidentification of its host galaxy or redshift would propagate into subsequent estimates of the local observed sGRB rate, but with varying levels of impact as the overall number of events in the sample has grown with time.

\subsection{Structured Jets}

In this work we adopt a top-hat jet geometry when relating the observed sGRB rate to the intrinsic event rate via the beaming factor $\eta = 1 - \cos\theta_{\rm j}$. While relativistic jets are likely structured, as demonstrated by the off-axis afterglow of GRB~170817A \citep{Troja2017,Lamb2017jet,Lazzati2018,Resmi2018,Mooley2018,DAvanzo2018,Alexander2018,Xie2018,Margutti2018,GG2018,Ghirlanda2019,Troja2020,McDowell:2023onw,Palmese:2023beh,Ryan2024}, the impact of such structure on the cosmological sGRB samples used to infer volumetric rates is expected to be limited \citep{Howell2025}. The empirical sGRB rate estimates employed in this work are derived from populations of bursts detected at cosmological distances, typically up to $z\sim1$ \citep{Coward2012,Wanderman2015} and in some studies extending to $z\sim2$ \citep{Ghirlanda2016,Salafia2023}. The rate estimates are thus dominated by events viewed on-axis close to the jet core \citep[e.g.,][]{OConnor2024MNRAS}, where the emission is brightest. Off-axis emission from structured jets becomes detectable only for nearby events \citep[e.g.,][]{Beniamini2019structuredjet,Howell2019,Colombo2022,Ronchini2022,Salafia2023,Howell2025,2026Kaur}, such as GRB~170817A \citep{Goldstein_2017,Savchenko_2017}, where the afterglow and prompt emission can be observed at viewing angles significantly outside the core \citep{Ioka2018,Kathirgamaraju2018,Lamb2017jet,Granot2017,BeniaminiNakar2019,Beniamini2019structuredjet}. At the redshifts characteristic of the bursts contributing to empirical sGRB rate estimates, the flux from such off-axis emission would fall below current detector thresholds, implying that most detected events are effectively probing the jet's core \citep{Salafia2019AandA,Salafia2023,OConnor2024MNRAS}. Consequently, treating $\theta_{\rm j}$ as an effective beaming angle is a reasonable approximation for interpreting cosmological sGRB samples. More detailed treatments that incorporate angular jet structure and integrate over viewing angle and luminosity functions would require additional assumptions about the jet structure and detection efficiency and cannot be straightforwardly applied to the empirically derived rates used here.

\subsection{Redshift evolution of the rate}

Although the observed sGRB sample is drawn primarily from cosmological redshifts \citep[e.g.,][]{2013Fong,2022Fong,OConnor2022}, its volumetric rate is often quoted as an equivalent local value at $z=0$ \citep[e.g.,][]{Coward2012}. In a compact binary merger scenario, however, the intrinsic sGRB rate density is not expected to be constant per unit comoving volume. Rather, it should evolve with redshift because the merger rate is set by the cosmic star formation history convolved with the delay time distribution between binary formation and coalescence. The rate at any epoch therefore depends on both the history of binary formation at earlier times and the fraction of those systems merging at that redshift.

This point is important because many empirical sGRB rate estimates are driven by bursts at moderate redshifts rather than by truly local events. For example, \citet{Coward2012} noted that the dominant contribution to their inferred $R_{\rm sGRB}$ comes from bursts at $z\sim0.2$--$0.4$, and that including redshift evolution could change the inferred local rate by roughly a factor of two, even though they did not apply such a correction because of uncertainties in the low redshift star formation history. Other population studies model this evolution explicitly \citep{Nakar2006,Wanderman2015,Ghirlanda2016,Salafia2023,pracchia_salafia2026}. These works convolve the cosmic star formation history with a delay time distribution to determine the observed redshift distribution of BNS mergers. A key difference between works is the inferred shape of the delay time distribution and the minimum delay time \citep[see, e.g.,][]{Ghirlanda2016,Beniamini2019,Zevin2022,pracchia_salafia2026}. Shorter delays (i.e., steeper delay time distributions) produce a merger rate density that declines more steeply toward $z=0$, whereas longer delays flatten the evolution, though this also depends on the exact assumed star formation history and redshift range.

As a result, the intrinsic sGRB or BNS merger rate can be significantly higher at the redshifts (median $z\approx 0.5$; \citealt{OConnor2022,2022Fong}) where the bulk of the sGRB population is observed than in the local Universe probed by LVK \citep{gwtc4,gwtc4pop}. For example, in the recent models favored by \citet{pracchia_salafia2026}\footnote{We note that \citet{pracchia_salafia2026} also find that there is no tension between their inferred BNS rate and current LVK estimates.}, the rate is higher by a factor of $\sim2$ at $z\sim0.5$ and by a factor of $\sim5$ at $z\sim1$ relative to $z=0$. The exact change depends sensitively on the assumed star formation rate and inferred delay time distribution. In any case, this means that a merger rate that appears only modest within the LVK horizon can still be fully consistent with the cosmological sGRB population observed at $z\sim0.5$--$2$. Properly accounting for redshift evolution, and in particular for the shorter delay times favored by recent works \citep{Beniamini2019,Zevin2022,Beniamini2024,pracchia_salafia2026,desantis2026}, can reduce the apparent tension between cosmological sGRB rate estimates and local LVK inferred BNS merger rate.

\section{Conclusion}
\label{sec:conclusions}

In this work, we investigate whether the on-axis isotropic local sGRB rates $R_{\rm sGRB}$ can be reconciled with current GW constraints on compact binary merger rates. We do this by combining the sGRB rate density estimates from the literature with the most up-to-date GW informed priors on the BNS and NSBH merger rate densities from the fourth LVK observing run. We infer the jet-opening angles and jet-launching fractions required to reproduce the observed sGRB population under a BNS-only progenitor scenario and a combined BNS+NSBH progenitor scenario. We also discuss how our conclusions depend on the adopted sGRB and GW rates, possible host redshift misidentification, bias in inferring the jet opening angle and on incorporating redshift evolution into the analysis. We also look at the possibility of incorporating other progenitor channels and the impact of adopted jet structure on our analysis. We summarize our findings as following:

\begin{itemize}
    \item Reconciling higher local sGRB rate estimates ($\sim8~\mathrm{Gpc^{-3}\,yr^{-1}}$, as inferred or assumed in the literature, e.g.,  \citealt{Coward2012,Fong2015ApJ815102,Sarin2022PRD,2023ApJ...959...13R}) with current LVK BNS merger rate constraints requires implausibly wide jet opening angles under reasonable jet-launching efficiencies. 

    \item For lower observed short GRB rate densities (described by a Poisson distribution with mean $\lambda_{\rm sGRB}\sim1-3~\mathrm{Gpc^{-3}\,yr^{-1}}$), the inferred BNS merger rate matches the rates inferred from LVK GWTC-4 catalog well. These sGRB rates are consistent with, e.g., \citet{Ghirlanda2016,Salafia2023}, as well as with the \citet{Coward2012} estimate that excludes GRB 080905, which has an uncertain host association. In this regime, the lower sGRB rates resolve the tension between jet opening angle constraints and current GW estimates, as it removes the need for wide jets and high intrinsic BNS merger rates. Consequently, the observed sGRB rates can be reconciled with the GW rates without invoking additional sGRB progenitor channels.

    \item Including NSBH mergers does not significantly modify the BNS jet properties needed to reconcile the sGRB and GW rates. Given their current volumetric rate and the modest jet-launching efficiencies predicted by our GW simulations ($f_{s,\rm NSBH}\lesssim0.23$), the NSBH channel contributes  a relatively small fraction (a median of $2-16\%$) of the observed sGRB population, compared to BNS mergers that contribute a median of $\sim 84-98\%$, depending on the assumed $R_{\rm sGRB}$. While future detections, including lower mass-gap NSBH systems, may provide important multimessenger constraints, NSBH mergers alone cannot reconcile high sGRB rate estimates with the current BNS merger rate constraints. Increasing $f_{s,\rm NSBH}$ could slightly enhance their contribution, but such high efficiencies are not supported by our GW simulations.

    \item Additional compact object merger channels, such as WDBH and WDNS systems, similarly do not resolve the tension between higher  sGRB rates (e.g., $\lambda_{\rm sGRB} = 8~\mathrm{Gpc^{-3}\,yr^{-1}}$) and the narrow jet opening angles inferred from afterglow observations. For the adopted volumetric rates, these channels would require either unrealistically high jet-launching efficiencies or extremely wide jets to contribute significantly to the observed sGRB rate.

    \item Commonly used distributions of sGRB jet opening angles, which are dominated by bursts with measured jet breaks may underestimate the prevalence of wide jets in the full sGRB population. If the true sGRB population includes a substantial fraction of wide jets, the current 90\% credible interval GW BNS rate constraints can accommodate observed sGRB rates up to $R_{\rm sGRB} < 7\,\mathrm{Gpc^{-3}\,yr^{-1}}$.
    
    \item If bursts with measured jet break detections are assumed to be representative of the full sGRB population, the inferred beaming correction is large, implying an intrinsic BNS merger rate substantially higher than current LVK BNS estimates. In this case, the sGRB rate allowed by the 90\% credible interval GW BNS constraints would be $R_{\rm sGRB} \lesssim 3.5~\mathrm{Gpc^{-3}\,yr^{-1}}$.

\end{itemize}

Looking ahead, joint GW-EM observations in future observing runs will provide improved constraints on the jet opening angle distribution and the contribution from different progenitor channels. 

\begin{acknowledgments}

The authors acknowledge useful discussions with Paz Beniamini, Simone Dichiara, Tim Dietrich, Maya Fishbach, Michael Moss, Rosalba Perna, and Geoff Ryan. A. P. is supported by NSF Grant No. 2308193. B. O. acknowledges support from the McWilliams Fellowship in the McWilliams Center for Cosmology and Astrophysics at Carnegie Mellon University. This research used resources of the National Energy Research Scientific Computing Center, a DOE Office of Science User Facility supported by the Office of Science of the U.S. Department of Energy under Contract No. DE-AC02-05CH11231 using NERSC award HEP-ERCAP0029208 and HEP-ERCAP0022871. This work used resources on the Vera Cluster at the Pittsburgh Supercomputing Center.

\end{acknowledgments}

\bibliography{References}{}
\bibliographystyle{aasjournal}
\end{document}